\documentclass[bibyear]{aa}  
\usepackage{longtable} 
\usepackage{pdflscape} 
\usepackage{multicol}
\usepackage{supertabular,booktabs}
\usepackage[normalem]{ulem}
\usepackage{amsmath}

\usepackage{placeins}
\usepackage{graphicx}
\usepackage[flushleft]{threeparttable}
\usepackage{txfonts}
\usepackage{diagbox}
\makeatletter
\renewcommand*\aa@pageof{, page \thepage{} of \pageref*{LastPage}}
\makeatother
\usepackage{natbib,twoopt}
\usepackage[breaklinks=true,linkcolor=blue,allcolors=blue,colorlinks=true]{hyperref} 
\bibpunct{(}{)}{;}{a}{}{,}             
\makeatletter
  \newcommandtwoopt{\citeads}[3][][]{\href{http://adsabs.harvard.edu/abs/#3}%
    {\def\hyper@linkstart##1##2{}%
     \let\hyper@linkend\@empty\citealp[#1][#2]{#3}}}
  \newcommandtwoopt{\citepads}[3][][]{\href{http://adsabs.harvard.edu/abs/#3}%
    {\def\hyper@linkstart##1##2{}%
     \let\hyper@linkend\@empty\citep[#1][#2]{#3}}}
  \newcommandtwoopt{\citetads}[3][][]{\href{http://adsabs.harvard.edu/abs/#3}%
    {\def\hyper@linkstart##1##2{}%
     \let\hyper@linkend\@empty\citet[#1][#2]{#3}}}
  \newcommandtwoopt{\citeyearads}[3][][]%
    {\href{http://adsabs.harvard.edu/abs/#3}
    {\def\hyper@linkstart##1##2{}%
     \let\hyper@linkend\@empty\citeyear[#1][#2]{#3}}}
\makeatother
\usepackage{orcidlink}

\usepackage{booktabs}
\usepackage{colortbl}

\begin{document} 

  \title{Blueberry galaxies up to 200~Mpc and their optical and infrared properties}
   \titlerunning{Blueberries up to 200 Mpc}

   \author{K. Kouroumpatzakis
          \inst{1}\,\orcidlink{0000-0002-1444-2016}\fnmsep\thanks{email:konstantinos.kouroumpatzakis@asu.cas.cz}
          \and 
          J. Svoboda\inst{1}\,\orcidlink{0000-0003-2931-0742}
          \and
          A. Zezas\inst{2,}\inst{3,}\inst{4}\,\orcidlink{0000-0001-8952-676X}
          \and 
          A. Borkar\inst{1}\,\orcidlink{0000-0002-9807-4520}
          \and
          E. Kyritsis\inst{2,}\inst{3}\,\orcidlink{0000-0003-1497-1134}
          \and 
          P. G. Boorman \inst{5}\,\orcidlink{0000-0001-9379-4716}
          \and 
          C. Daoutis\inst{2,}\inst{3,}\inst{4}\,\orcidlink{0000-0002-3028-4759}
          \and
          B. Adamcová\inst{1}\,\orcidlink{0009-0008-6899-4749}
          \and 
          R. Grossová\inst{1,}\inst{6}\,\orcidlink{0000-0003-3471-7459}
          }

   \institute{
             Astronomical Institute, Academy of Sciences, Boční II 1401, CZ-14131 Prague, Czech Republic\\
              \email{konstantinos.kouroumpatzakis@asu.cas.cz}
             \and
             Department of Physics, University of Crete, Voutes University campus, 70013 Heraklion, Greece
            \and
              Institute of Astrophysics, Foundation for Research and Technology-Hellas, N. Plastira 100, Vassilika Vouton, 71110 Heraklion, Greece
             \and
             Center for Astrophysics \textbar\ Harvard \& Smithsonian, 60 Garden St., Cambridge, MA 02138, USA
             \and
             Cahill Center for Astronomy and Astrophysics, California Institute of Technology, Pasadena, CA 91125, USA
            \and
             Department of Theoretical Physics and Astrophysics, Faculty of Science, Masaryk University, Kotlářská 2, Brno, CZ-61137, Czech Republic.
             }

   \date{Received ; accepted }

 
  \abstract
   {
   Dwarf highly star-forming galaxies (SFGs) dominated the early Universe and are considered the main driver of its reionization. 
   However, direct observations of these distant galaxies are mainly confined to rest-frame ultraviolet and visible light, limiting our understanding of their complete properties. 
   Therefore, it is still paramount to study their local analogs, the green pea (GP) and blueberry (BB) galaxies. 
   }
   {
   This work aims to expand our knowledge of BBs by identifying a new sample that is closer and in the southern sky. 
   In addition to the already known BBs, this new sample will allow for a statistically significant study of their properties probed by visible and infrared (IR) light.
   }
   {
   By utilizing the HECATE catalog, which provides optical and IR photometry and characterization of galaxies, along with data from Pan-STARSS and SDSS, this study selects and analyzes a new sample of BBs. 
   We employed spectral energy distribution fitting to derive homogeneous measurements of star-formation rates and stellar masses. 
   Additionally, we measured emission-line fluxes, including $\rm \ion{He}{II}~\lambda 4686$, through spectral fitting.
   }
   {
    Through this work, we identified 48 BBs, of which 40 were first recognized as such, with the nearest at 19~Mpc.
    Fourteen of the BBs are in the south sky. 
    The BBs tend to be extremely IR red in both WISE $W1-W2$ and $W2-W3$ colors, distinguishing them from typical SFGs. 
    Dwarf SFGs with higher specific star-formation rates tend to have redder IR colors. 
   }
   {
   Blueberry galaxies stand out as the most intensely star-forming sources in the local Universe among dwarf galaxies. 
   They exhibit unique characteristics, such as being intrinsically bluer in visible light, redder in the infrared, and less massive.
   They also have higher specific star-formation rates, equivalent widths, lower metallicities, and the most strongly ionized interstellar medium compared to typical SFGs and GPs. 
   }


   \maketitle
%
\section{Introduction}
\label{sec:intro}

After the Dark Ages, the early Universe transformed from being mostly opaque to being transparent due to the reionization of its intergalactic medium (IGM) during the reionization era \citep[see e.g.,][ and references therein]{2001PhR...349..125B}, which took place at $6 \lesssim z \lesssim 20$ \citep[e.g.,][]{2015ApJ...811..140B,2015ApJ...802L..19R}.
What caused this pivotal change and which sources could have produced the copious amounts of the required ionizing radiation are still open questions.
Dwarf highly star-forming galaxies (SFGs) are mainly considered to be responsible because they were numerous and capable of producing large amounts of ultraviolet (UV) and Lyman continuum (LyC) radiation due to hosting a large stellar population (SP) of massive stars \citep[e.g.,][]{1987ApJ...321L.107S,2001ARA&A..39...19L}.
Active galactic nuclei (AGNs) and quasars could have also contributed substantially to the reionization of the IGM \citep[e.g.,][]{2009ApJ...703.2113V}.  

Recently, the James Webb Space Telescope (JWST) revolutionized our view of the young Universe by allowing the direct study of the rest-frame UV and visible light of some very distant galaxies at redshifts $z>7$ 
\citep[e.g.,][]{2023ApJS..265....5H,2023Natur.616..266L}.
However, mainly due to their immense distances, the direct study of these early galaxies through other wavelengths (e.g., X-rays, radio) is beyond our current capabilities. 
Thus, their local analogs, namely green pea (GP) and blueberry (BB) galaxies, offer the only alternative solution.
In fact, recent JWST observations of high-redshift star-forming dwarf galaxies revealed that they show a remarkable resemblance to known and relatively nearby GPs \citep[][]{2022A&A...665L...4S,2023ApJ...942L..14R}.

Green peas are dwarf starburst galaxies relatively nearby at a redshift of $z \sim 0.25$ \citep[][]{2009MNRAS.399.1191C}.
They are characterized by low stellar masses ($M_\star \simeq 10^9 ~ M_\odot$), compact scales ($\rm \sim ~ 5 \, kpc $), sub-solar metallicities (log(O/H)+12~$\simeq$~8.1), and high star-formation rates (SFRs; i.e.,~$ > 10 ~ M_\odot ~ {\rm yr}^{-1}$).
Their distinct green color is mainly due to strong $\rm [\ion{O}{III}] \lambda \lambda 4363, 5007$ emission lines.
Moreover, GPs show significant LyC radiation escape fractions \citep[$>10\%$;][]{2013ApJ...766...91J,2016Natur.529..178I,2016MNRAS.461.3683I} which is not common in the low-redshift Universe.
Blueberry galaxies are also highly star-forming dwarfs, but they are smaller, have even lower metallicities (log(O/H)+12~$<7.8$), are closer ($z<0.05$), and are considered smaller counterparts to GPs \citep[][]{2017ApJ...847...38Y}.

Most homogeneous samples of GPs and BBs \citep[][]{2011ApJ...728..161I,2019ApJ...874...52M,2019ApJ...885...96J,2020ApJ...898...68B,2022ApJ...927...57L,2023AJ....166..133D} are based on the Sloan Digital Sky Survey (SDSS).
The sample of \cite{2017ApJ...847...38Y} was based on SDSS photometry, while the rest took advantage of the spectroscopic catalogs available by SDSS.
The work of \cite{2022ApJ...927...57L} was based on the spectroscopic coverage of the Large Sky Area Multi-Object Fibre Spectroscopic Telescope \citep[LAMOST;][]{1996ApOpt..35.5155W,2004ChJAA...4....1S}, and \cite{2022AJ....163..150K} has also identified some GPs with the help of Hubble Space Telescope (HST) archival data.
Overall, the vast majority of known GP and BB samples are limited to the north part of the equatorial sky and relatively large distances.

Apart from the properties of these sources that are known through the visible spectrum, only a few works have studied, for instance, their X-ray \citep[][]{2019ApJ...880..144S} or radio \citep{2019ApJ...882L..19S,2024arXiv240319635B} emission, but these studies were limited to a small number of sources.
Near-infrared observations were used by \cite{2003A&A...410..481N} to study the photometric structure of nearby compact blue dwarf galaxies (BDGs). 
The WISE survey \citep{2010AJ....140.1868W} has also been used for the discovery of BDGs. 
\cite{2011ApJ...736L..22G} discovered two sources with extremely red colors in mid-IR wavelengths.
\cite{chandola2023h} selected 11 blue compact galaxies with bright mid-IR emission and studied their \ion{H}{I} emission, and \cite{2022A&A...668A..60L} found 466 extreme emission-line galaxies combining the Javalambre Photometric Local Universe Survey \citep[J-PLUS;][]{2019A&A...622A.176C} and the ALLWISE catalog \citep{2014yCat.2328....0C}. 

The majority of the previous studies have focused on high-z samples, but local samples are needed to be able to study the properties of BBs in more detail and for multi-wavelength follow up.
An extended and relatively more nearby sample of BBs will allow for a statistically significant study of the BBs' IR and visible emission and could also serve as a base for future observations and further investigation.
This work aims to expand the sample of one of the most extreme classes of SFGs, the BBs, using the Heraklion Extragalactic Catalogue \citep[HECATE;][]{2021MNRAS.506.1896K}.\footnote{\url{https://hecate.ia.forth.gr/}}
This work is organized as follows: Section \ref{sec:criteria} describes the selection criteria, and Section \ref{sec:results} shows the resulting sample, the main results regarding their visible and IR properties, and comparisons with other samples of BDGs, GPs, and BBs. 
A discussion based on the results is presented in Section \ref{sec:Discussion}, and Section \ref{sec:conclusions} concludes the article.
A \cite{2016A&A...594A..13P} cosmology is adopted throughout the paper ($\rm \Omega_m=0.308$, $h=0.678$).
Unless stated otherwise, uncertainties and confidence intervals (CIs) refer to 68\%, and the size of the histogram's bins is based on Scott's rule \citep{10.1093/biomet/66.3.605}.

\section{Selection criteria}
\label{sec:criteria}

In order to extend the sample of BBs beyond the footprint of SDSS, we performed a search that depended on the photometric rather than the spectroscopic properties of the sources.
The selection was based on HECATE, which is an all-sky catalog of 204,733 galaxies within redshift $\rm z \leq 0.047$ ($\rm D \leq 200$~ Mpc) with more than 50\% completeness in B-band luminosity up to 170~Mpc.
It offers distances, sizes, positions, SFRs, stellar masses, and metallicities as well as morphological and nuclear activity classifications along with the galaxies' IR and optical photometry.
Thus, it is perfectly suited to selecting specific types of galaxies and examining their properties. 
We performed our search using the second enhanced version of HECATE, HECATEv2.0 (Kyritsis et~al. in prep), which provides additional information, such as optical photometry from Pan-STARRS \citep{2016arXiv161205560C}, and more robust estimations of stellar masses and SFRs that were calculated using the methods of \cite{2023A&A...673A..16K}.
HECATEv2.0 provides photometry through Pan-STARRS for the majority of the sources covered by this survey down to a declination of $-30^{\circ}$.

We adopted the Pan-STARRS \texttt{Kron} and the point-spread function (PSF) photometry, which was also used for the selection since most BBs appear to be spatially compact \citep{2017ApJ...847...38Y}.
We also adopted the SDSS \texttt{cModelMag} and the \texttt{Petrosian} photometries provided by SDSS. 
To avoid missing BBs, which have some of the bluest colors found in galaxies, all four aforementioned photometric systems were used in a complementary manner.
This allowed us to cross-test the results since on many occasions one of the different photometries could have varied for different reasons (e.g., contamination from other sources, effects such as varying photometric or atmospheric conditions between the separate filter observations). 

\begin{figure}
    \centering
    \includegraphics[width=\columnwidth]{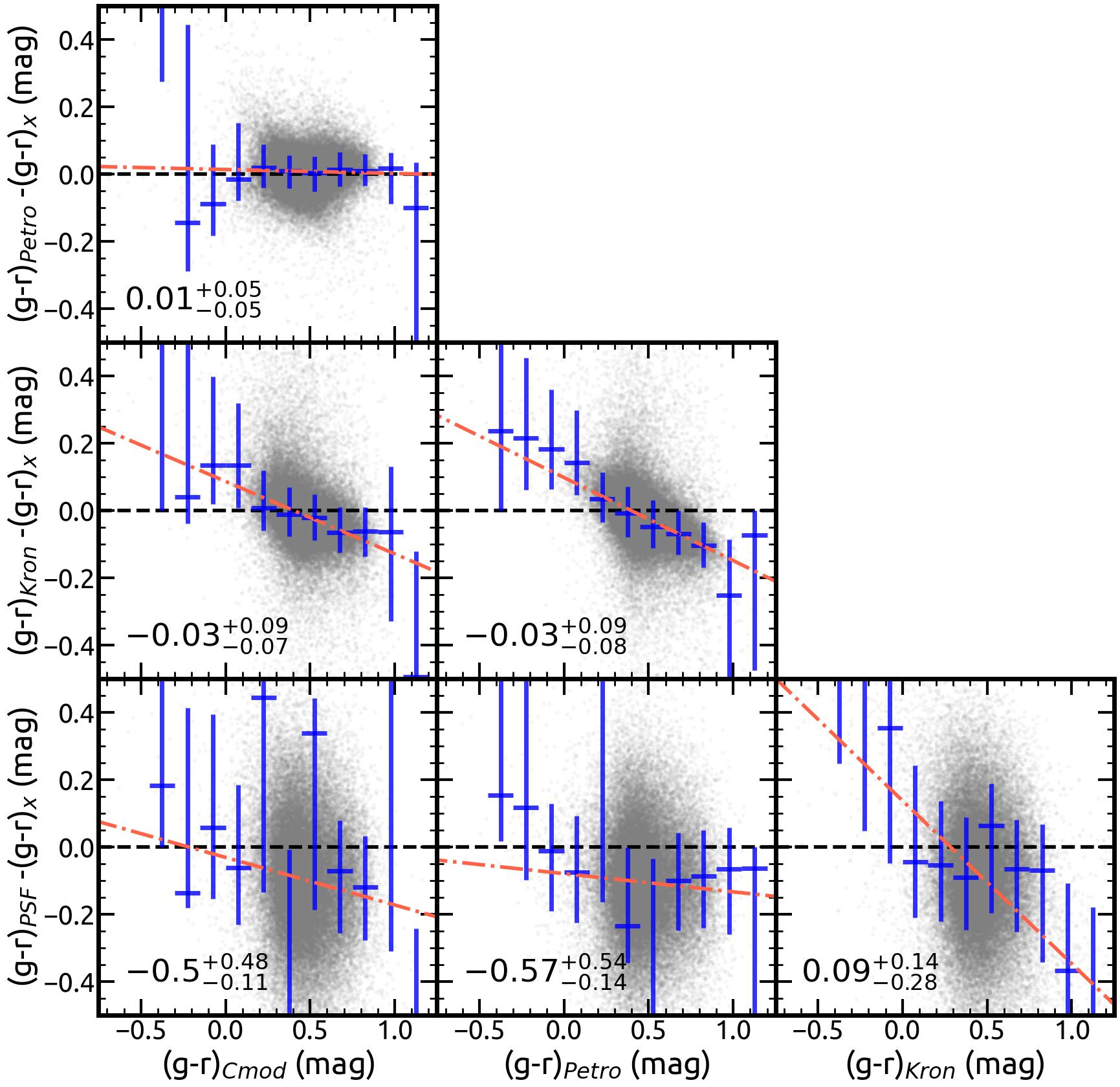}
    \caption{Comparison of the $g-r$ colors of the SDSS \texttt{cModelMag}, \texttt{Petrosian}, and Pan-STARRS \texttt{Kron} and \texttt{PSF} photometries for SFGs (no AGNs, LINER, or composite) of HECATE. 
    The abscissa shows the $g-r$ color, and the ordinate shows the difference of the $g-r$ colors of the photometric schemes as mentioned by each panel's axis. 
    The blue error bars represent the modes and 68\% CIs of the distribution of galaxies in $g-r$ bins with a size of 0.15 mag.
    The red dashed-dotted line represents their linear regression fit.
    At the bottom left of each panel is the mode and 68\% CI of the distribution of the color difference accounting for all sources.
    The black dashed line represents equality.}
    \label{fig:HEC_opt_photometries}
\end{figure}

In order to quantify their differences and define the final photometric scheme for the selection of the BBs, we compared the $g-r$ colors of the optical photometric schemes mentioned above (Figure \ref{fig:HEC_opt_photometries}).
This comparison involved all the galaxies classified as star forming in HECATE (no AGNs, composite, or LINERs) with a signal-to-noise ratio higher than three (S/N~$>3$) in the g and r bands in all four Pan-STARRS and SDSS photometries.
The best agreement was found between the \texttt{cModelMag} and \texttt{Petrosian} magnitudes, as SDSS provides both.
The standard deviation of the difference between their $g-r$ colors is 0.05 dex.
The Pan-STARRS \texttt{Kron} magnitudes show good agreement on average, $\langle (g-r)_{\rm Kron} - (g-r)_{\rm cModel} \rangle = -0.03$, and the standard deviation of their difference is $\simeq 0.08$ dex. 
However, galaxies based on Pan-STARRS \texttt{Kron} magnitudes appear to be redder for blue galaxies and bluer for red ones concerning both \texttt{cModelMag} and \texttt{Petrosian} SDSS photometries. 
As expected, the Pan-STARRS \texttt{PSF} photometry shows the largest scatter compared to the rest of the photometries.
Moreover, it shows, on average, $-0.1$ magnitude bluer colors compared to Kron photometry throughout the color range.
The results of the linear regression fits between the $g-r$ colors of the different photometries are summarized in Table \ref{tab:Linear_Phots}.

\begin{table}[h!]
    \centering
    \begin{tabular}{c|ccc}
        & ($\alpha, \beta$) & ($\alpha, \beta$) & ($\alpha, \beta$) \\
        \hline
        \texttt{Petrosian} & 0.02, 0.99 \\
        \texttt{Kron} & 0.09, 0.78 & 0.1, 0.76\\
        \texttt{PSF} & -0.03, 0.86 & -0.08, 0.94 & 0.07, 0.64\\
        \hline
                     & \texttt{cModelMag} & \texttt{Petrosian} & \texttt{Kron}
    \end{tabular}
    \caption{Results of the linear regression fits between the $g-r$ colors of the different photometries.
    The fits are in the form of $(g-r)_{y} = \alpha + \beta (g-r)_{x}$.}
    \label{tab:Linear_Phots}
\end{table}

The \texttt{cModelMag} photometry corresponds to an aperture defined by a radial profile that is a linear combination of the best fit of an exponential and a de Vaucouleurs profile.
Thus, because \texttt{cModelMag} is defined to encompass most of the light of extended sources (e.g., galaxies), it is considered the most trustworthy photometric scheme for this work, and therefore, we used it to define the selection criteria. 
However, to be able to expand the BB sample to parts of the sky not covered by SDSS, we also took into account sources that were measured only by Pan-STARRS.

The selection of the galaxies was based on their photometric properties following the scheme proposed by \cite{2017ApJ...847...38Y} who, in order to find BBs, selected sources with $g-r < -0.5$, $r-i < 1$, $g-u < -0.3$, and $g-i < -0.5$~(mag), adopting the SDSS \texttt{cModelMag} photometry.
These criteria were defined by simulating the emission of BBs with intensive star formation and strong ionization that lead to large equivalent widths ($\rm EW_{[\ion{O}{III}] \lambda 5007} > 800$~\AA) and very young SP ages.
The simulations modeled galaxies with an old SP component with an age equal to 900~Myr and one very recent starburst episode at 4~Myr with the code \texttt{Starburst99} \citep{1999ApJS..123....3L}.

For this work, we allowed slightly more relaxed criteria of $+0.1$ mag, which corresponds to the scatter of the comparison between the \texttt{cModelMag} and \texttt{Kron} photometries (Table \ref{tab:Linear_Phots}) and partially compensates for photometric uncertainties or slightly different star formation histories (SFHs) and metallicities from those assumed by \cite{2017ApJ...847...38Y}.
Thus, the selection limit for the SDSS \texttt{cModelMag} and \texttt{Petrosian} photometries is $g-r=-0.4$~(mag).
Also, because Pan-STARRS \texttt{Kron} photometry appears redder for blue galaxies compared to SDSS (Figure \ref{fig:HEC_opt_photometries}), we further allowed for $+0.15$~mag for the selection criteria applied to Pan-STARRS photometry.
This value was estimated based on the linear regression fit (Table \ref{tab:Linear_Phots}) between SDSS \texttt{cModelMag} and Pan-STARRS \texttt{Kron} photometries applied to the limit $g-r=-0.4$~(mag).
We kept galaxies that fulfill these criteria simultaneously in the SDSS \texttt{cModelMag} and \texttt{Petrosian} photometries or simultaneously in the Pan-STARRS \texttt{Kron} and \texttt{PSF} photometries.
Additionally, we calculated the average colors by combining all the mentioned photometries.
Sources with average colors that did not fulfill the criteria applied to Pan-STARRS photometry were removed (three sources were omitted).
We also omitted sources with problematic photometry by limiting the accepted colors to $>-2$ and respective uncertainties $\rm < 0.2 ~ (mag)$.
The selection criteria are summarized in Equation \ref{eq:Selection} shown in the Appendix \ref{sec:Appendix_A}.
This filtering process resulted in 83 sources that fulfilled the selection criteria.

Moreover, because we are interested in the IR properties of the galaxies, we omitted sources where the WISE extended source catalog flag \texttt{star\_contamination} is true in order to avoid sources with photometries contaminated by bright stars.
This was the case for eight sources (PGC~360, 5817, 6336, 3876, 42031, 45482, 8881, and 4094118), all of which are spectroscopically confirmed stars and were removed.
The final step involved visual inspection of the selected galaxies.
The galaxies that were omitted through visual inspection are 
PGC~46329, 48755, 4081747, 46495, 4084118, and 6723779, which are also spectroscopically confirmed stars.
The reasons for their omission are as follows: PGC~5060407 due to its presence in a crowded field of background and foreground sources; PGC~28757, 6741333, 6741249, 4107113, 6742056, 5059213, 6740710, 40683, and 5057032 because they resemble ultra-faint galaxy candidates; PGC~5078149, 774317, 1076594, and 101538 because they are blue star-forming regions that are part of significantly larger or spiral galaxies; and PGC~33553, 53905, 49767, 56870, 41771, 5082584, and 46982, as they are sources that have strongly contradicting photometry between SDSS and Pan-STARRS.

\section{Results}
\label{sec:results}

\subsection{Proposed sample of blueberry galaxies}
\label{sec:Sample_of_BBs}

\begin{figure*}
    \centering
    \includegraphics[width=\textwidth]{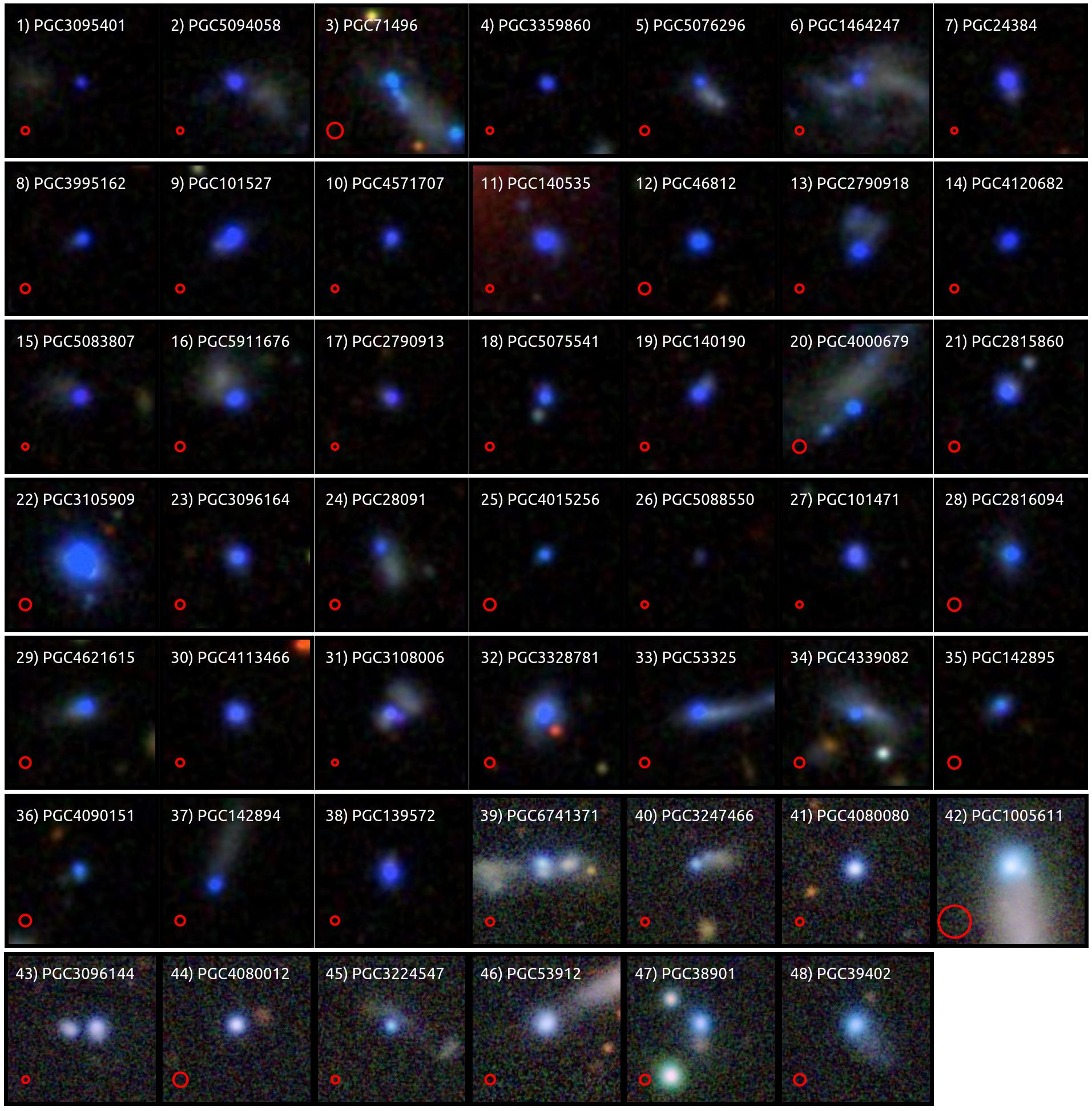}
    \caption{Optical images of the g, r, and i bands of the BBs of this work. 
    Pictures of BBs~1 up to BB~38 are as imaged by SDSS, while the rest are by Pan-STARRS.
    All images are in $25.6'' \times 25.6''$ frames except BB~42, which is shown in a $51'' \times 51''$ frame.
    The red circle at the bottom left of each image corresponds to a diameter equal to 1~kpc at the respective distance of each galaxy.
    }
    \label{fig:opt_images}
\end{figure*}

The selection process yielded 48 BBs, with the most distant at 198~Mpc and the nearest at 19~Mpc.
Table \ref{tab:Blueberries_tab1} reports the selected galaxies and their $g-r$ and $r-i$ colors from the different photometries.
The sources are reported in ascending order of their weighted $g-r$ color, combining all photometries with weights equal to 1.2, 1.1, 1.0, and 0.9 for the SDSS \texttt{cModelMag}, SDSS \texttt{Petrosian}, Pan-STARRS \texttt{Kron}, and Pan-STARRS \texttt{PSF}, respectively.
Table \ref{tab:Blueberries_tab2} reports some fundamental characteristics of the galaxies, including their $M_\star$, specific SFR (sSFR; $\rm sSFR \equiv SFR/M_\star$), sizes, and distances along with metallicities and EWs of the $\rm [\ion{O}{II}]\lambda3726$, $\rm [\ion{O}{III}]\lambda5007$, H$\beta$, and H$\alpha$ emission lines for sources that have spectra and were analyzed by MPA-JHU \citep{10.1111/j.1365-2966.2003.07154.x,10.1111/j.1365-2966.2004.07881.x,Tremonti_2004}.
Uncertainties for Tables \ref{tab:Blueberries_tab1} and \ref{tab:Blueberries_tab2} are provided in the online\footnote{Tables \ref{tab:Blueberries_tab1} and \ref{tab:Blueberries_tab2} are available in electronic form at the CDS via anonymous ftp to cdsarc.u-strasbg.fr (147.231.47.162) or via \url{http://cdsweb.u-strasbg.fr/cgi-bin/qcat?J/A+A/}}
version.
Figure \ref{fig:opt_images} shows the optical images composed of g, r, and i band observations of the selected sources as imaged by SDSS or Pan-STARRS, for sources not covered by SDSS.

The nearest BB found is BB~42 (PGC~1005611) at 19~Mpc. 
The second nearest source is BB~3 (PGC~71496) at 75~Mpc, and the third nearest is BB~44 (PGC~4080012) at 86~Mpc.
There are eight sources in common with the \cite{2017ApJ...847...38Y} BBs.
These are the BBs PGC 3359860, 3995162, 101527, 4571707, 46812, 139572, 4120682, and 3328781, and they are also part of the \cite{2019ApJ...874...52M} and \cite{2019ApJ...885...96J} samples. 
In the following analysis, for comparisons that include \cite{2017ApJ...847...38Y} BBs, we only consider new sources identified by this work.

Although \cite{2017ApJ...847...38Y} reports 25 BBs at distances below 200 Mpc ($z \approx 0.0437$), 17 of them are not included in HECATE.
However, 15 of these 25 BBs have distances higher than 175~Mpc, where HECATE's completeness drops significantly.
The eight galaxies that are included in HECATE have been successfully recognized as BBs by our selection method.
The nearest BB reported by \cite{2017ApJ...847...38Y} is at 101 Mpc, while this work reports seven galaxies below 100 Mpc.
The mode of the distribution of distances of \cite{2017ApJ...847...38Y} BBs is $184^{+43}_{-35}$~Mpc, while the mode of the distribution of distances of BBs in this work is $123^{+48}_{-28}$~Mpc.

\begin{table*}[ht!]
    \setlength{\tabcolsep}{5.5pt} 
    \centering
    \caption{Observed $g-r$ and $r-i$ colors of the BBs of this work from the different photometries. 
    The sources are arranged in ascending order based on their weighted average $g-r$ color. 
    }
    \begin{threeparttable}
    \begin{tabular}{clcc|c|cccc|cccc}
    ID & PGC & R.A. & Dec. & $r$ & $g{-}r$ & $g{-}r$ & $g{-}r$ & $g{-}r$ & $r{-}i$ & $r{-}i$ & $r{-}i$ & $r{-}i$ \\
     & & & & * & \texttt{cMod} & \texttt{Petro} & \texttt{Kron} & \texttt{PSF} & \texttt{cMod} & \texttt{Petro} & \texttt{Kron} & \texttt{PSF}\\
     &  & (Deg.) & (Deg.) & (mag) & (mag) & (mag) & (mag) & (mag) & (mag) & (mag) & (mag) & (mag)\\
     \hline
    1 & 3095401 & 2.4156 & 2.5718 & 21.3 & -1.4 & -1.4 & -0.8 & -0.6 & 0.1 & 0.1 & -0.8 & -0.9 \\
    2 & 5094058 & 33.277565 & 0.936803 & 18.8 & -1.2 & -1.2 & -0.4 & -0.6 & 0.3 & 0.2 & -0.2 & -0.6 \\
    3 & 71496 & 351.935088 & -2.013199 & 18.7 & -0.7 & -0.7 & -1.2 & -0.5 & -1.1 & -1.0 & -0.2 & -0.6 \\
    4 & 3359860$^\dagger$ & 221.172456 & 4.16165 & 20.1 & -0.9 & -0.9 & -0.5 & -0.4 & 0.0 & 0.0 & -0.5 & -0.6 \\
    5 & 5076296 & 48.250193 & 0.103381 & 20.1 & -0.7 & -0.9 &  & -1.3 & -0.1 & -0.4 &  & -1.0 \\
    6 & 1464247 & 229.631954 & 14.677362 & 17.2 & -1.1 & -1.1 & 0.0 & -0.6 & 0.1 & 0.1 & -0.1 & -0.5 \\
    7 & 24384 & 130.12455 & 47.119511 & 18.0 & -0.9 & -0.9 & -0.3 & -0.4 & 0.2 & 0.2 & -0.4 & -0.5 \\
    8 & 3995162$^\dagger$ & 208.856988 & 46.86428 & 19.9 & -0.8 & -0.8 & -0.5 & -0.6 & -0.1 & -0.4 & -0.6 & -1.0 \\
    9 & 101527$^\dagger$ & 227.392233 & 37.529368 & 17.9 & -0.8 & -0.9 & -0.5 & -0.5 & -0.1 & -0.1 & -0.5 & -0.8 \\
    10 & 4571707$^\dagger$ & 126.418617 & 18.771452 & 19.6 & -0.8 & -0.9 & -0.5 & -0.3 & 0.0 & 0.0 & -0.4 & -0.6 \\
    11 & 140535 & 236.431458 & 8.967036 & 17.3 & -0.9 & -0.9 & -0.4 & -0.4 & 0.1 & 0.0 & -0.5 & -0.6 \\
    12 & 46812$^\dagger$ & 200.947742 & -1.54774 & 18.9 & -0.7 & -0.7 & -0.6 & -0.6 & -0.6 & -0.7 & -0.8 & -0.8 \\
    13 & 2790918 & 172.6671 & 43.7327 & 18.8 & -0.6 & -0.9 & -0.6 & -0.5 & -0.1 & -0.2 &  & -0.8 \\
    14 & 4120682$^\dagger$$^\ddagger$ & 242.043072 & 35.469386 & 19.8 & -1.1 & -1.2 &  &  & -0.2 & -0.2 &  &  \\
    15 & 5083807 & 145.684329 & 41.17204 & 18.9 & -0.4 & -0.9 & -0.4 & -0.7 & 0.0 & 0.5 & -0.2 & -0.3 \\
    16 & 5911676 & 217.452528 & -1.16935 & 18.1 & -0.8 & -0.8 & -0.2 & -0.5 & -0.3 & -0.3 & -0.4 & -0.6 \\
    17 & 2790913 & 168.48495 & 44.308899 & 19.2 & -0.6 & -0.8 & -0.4 & -0.5 & 0.3 & 0.4 & -0.1 & -0.3 \\
    18 & 5075541 & 345.541623 & 0.82752 & 19.1 & -0.6 & -0.7 & -0.3 & -0.5 & -0.2 & -0.2 & -0.1 & -0.5 \\
    19 & 140190 & 203.438757 & 28.753155 & 18.6 & -0.6 & -0.7 & -0.3 & -0.4 & 0.0 & 0.0 & -0.3 & -0.6 \\
    20 & 4000679 & 169.686258 & 2.908797 & 18.8 & -0.5 & -0.6 &  & -0.9 & -0.6 & -0.8 &  & -0.5 \\
    21 & 2815860 & 137.78565 & 54.180801 & 17.9 & -0.6 & -0.6 & -0.3 & -0.3 & -0.1 & -0.3 & -0.4 & -0.5 \\
    22 & 3105909 & 193.27488 & -3.216273 & 15.6 & -0.5 & -0.5 & -0.3 & -0.5 & -0.6 & -0.6 & -0.5 & -0.6 \\
    23 & 3096164 & 189.189748 & 26.301151 & 18.5 & -0.5 & -0.6 & -0.3 & -0.5 & 0.0 & -0.2 & -0.3 & -0.4 \\
    24 & 28091 & 146.69415 & 55.951801 & 19.5 & -0.5 & -0.6 & -0.2 & -0.4 & -0.1 & -0.2 & -0.8 & -0.4 \\
    25 & 4015256 & 121.991636 & 34.244234 & 19.7 & -0.4 & -0.5 & -0.3 & -0.4 & -0.6 & -0.6 & -0.4 & -0.7 \\
    26 & 5088550 & 23.835057 & 20.153192 & 21.6 & -0.4 & -0.4 & -0.4 & -0.4 & 0.3 & 0.1 & 0.0 & -0.3 \\
    27 & 101471 & 186.153029 & 37.41013 & 18.2 & -0.6 & -0.6 & -0.2 & -0.2 & 0.1 & 0.1 & -0.3 & -0.4 \\
    28 & 2816094 & 215.661924 & 54.235839 & 18.1 & -0.4 & -0.5 & -0.3 & -0.5 & -0.2 & -0.6 & -0.4 & -0.7 \\
    29 & 4621615 & 202.351482 & 22.61984 & 18.1 & -0.3 & -0.5 & -0.3 & -0.5 & -0.1 & -0.3 & -0.3 & -0.7 \\
    30 & 4113466 & 215.928689 & 22.958003 & 18.2 & -0.5 & -0.5 & -0.2 & -0.2 & 0.0 & -0.1 & -0.4 & -0.5 \\
    31 & 3108006 & 222.945728 & -0.945494 & 18.0 & -0.4 & -0.5 & -0.3 & -0.3 & 0.1 & 0.1 & 0.0 & -0.1 \\
    32 & 3328781$^\dagger$$^\ddagger$ & 222.022369 & -1.182752 & 16.8 & -0.4 & -0.5 & -0.2 & -0.4 & -0.2 & -0.2 & -0.3 & -0.5 \\
    33 & 53325 & 223.775214 & 38.137962 & 17.7 & -0.4 & -0.5 & -0.2 & -0.3 & -0.1 & -0.2 & -0.2 & -0.4 \\
    34 & 4339082 & 190.908092 & 17.933047 & 17.5 & -0.4 & -0.5 & -0.1 & -0.5 & -0.4 & -0.4 &  &  \\
    35 & 142895 & 185.254236 & 30.62877 & 19.1 & -0.2 & -0.4 & -0.3 & -0.5 & -0.2 & -0.4 & -0.2 & -0.5 \\
    36 & 4090151 & 145.72608 & 34.069969 & 19.5 & -0.2 & -0.4 & -0.3 & -0.4 & -0.4 & -0.4 & -0.3 & -0.6 \\
    37 & 142894 & 183.752147 & 29.918301 & 19.5 &  &  & -0.7 & -0.4 &  &  & -0.7 & -0.6 \\
    38 & 139572$^\dagger$ & 170.95341 & 20.841173 & 18.0 &  &  & -0.4 & -0.8 &  &  & -0.5 & -0.4 \\
    39 & 6741371 & 39.545832 & 33.939447 & 18.3 &  &  & -0.3 & -0.8 &  &  & -1.1 & -0.7 \\
    40 & 3247466 & 337.71915 & -23.9214 & 19.8 &  &  & -0.3 & -0.6 &  &  & -0.6 & -0.5 \\
    41 & 4080080 & 338.399053 & -26.470739 & 18.4 &  &  & -0.4 & -0.4 &  &  & -0.4 & -0.3 \\
    42 & 1005611 & 42.065712 & -8.290536 & 16.9 &  &  & -0.4 & -0.4 &  &  & -0.9 & -1.0 \\
    43 & 3096144 & 187.0746 & -11.5594 & 18.9 &  &  & -0.3 & -0.5 &  &  & -0.1 & -0.2 \\
    44 & 4080012 & 36.612792 & -11.416837 & 18.8 &  &  & -0.4 & -0.4 &  &  & -0.3 & -0.3 \\
    45 & 3224547 & 339.259787 & -28.878851 & 19.4 &  &  & -0.3 & -0.4 &  &  & -0.2 & -0.9 \\
    46 & 53912 & 226.5405 & 39.178501 & 18.3 &  &  & -0.3 & -0.4 &  &  & -0.4 & -0.5 \\
    47 & 38901 & 183.082966 & -28.813061 & 18.4 &  &  & -0.3 & -0.4 &  &  & -0.7 & -0.8 \\
    48 & 39402 & 184.321209 & -28.042389 & 18.2 &  &  & -0.4 & -0.3 &  &  & -0.8 & -1.1 \\
    \end{tabular}
    \begin{tablenotes}
      \small
      \item * SDSS \texttt{cModelMag} or Pan-STARRS \texttt{Kron} when \texttt{cModelMag} is not available.
      \item $^\dagger$ denotes BBs in common with \cite{2017ApJ...847...38Y}.
      $^\ddagger$ denotes BBs in common with \cite{2019ApJ...874...52M} and \cite{2019ApJ...885...96J}.
      \item Uncertainties are provided in the electronic form of the table.
      \end{tablenotes}
  \end{threeparttable}
    \label{tab:Blueberries_tab1}
\end{table*}

\begin{table*}
    \setlength{\tabcolsep}{5.3pt} 
    \centering
    \caption{Redshift, distances, sizes, stellar masses, sSFRs, metallicities, EWs, and $\rm \ion{He}{II} \lambda 4684$ fluxes of the proposed BBs. }
    \begin{threeparttable}
    \begin{tabular}{c|ccccccccccccc}
    ID & redshift. & Dist. & size & log~$M_\star$ & log sSFR & metallicity & $\rm EW_{\ion{O}{II}}$ & $\rm EW_{H\beta}$ & $\rm EW_{\ion{O}{III}}$ & $\rm EW_{H\alpha}$ &  log~$ f_{\rm \ion{He}{II}}$ & Compact \\
     & MPA & HEC/S & HEC & SED & SED & MPA & MPA & MPA & MPA & MPA & FADO &\\
     & & (Mpc) & (kpc) & $(M_\odot)$ & $\rm (yr^{-1})$ & (12+log(O/H)) & (\AA) & (\AA) & (\AA) & (\AA) & $\rm (erg/s/cm^2)$ & \\
     \hline
1 &  & 157 &  & 6.9 & -7.4 &  &  &  &  &  &  & N \\
2 & 0.0399 & 182 & 1.9 & 7.0 & -7.0 & 8.6 & 89 & 191 & 1194 & 1000 & -15.8 & N \\
3 &  & 75 & 3.9 & 6.4 & -6.8 &  &  &  &  &  &  & N \\
4 & 0.0388 & 176 & 2.0 & 7.3 & -7.6 & 7.7 & 54 & 183 & 1186 & 841 & -16.2 & Y \\
5 & 0.0292 & 131 & 4.1 & 5.9 & -6.3 & 7.9 & 31 & 125 & 711 & 577 &  & N \\
6 &  & 151 & 8.4 & 7.0 & -7.2 &  &  &  &  &  &  & N \\
7 & 0.0421 & 192 & 5.1 & 7.6 & -6.4 & 7.8 & 49 & 202 & 1143 & 1028 & -15.3 & Y \\
8 & 0.0281 & 127 & 2.5 & 7.2 & -7.7 &  & 32 & 230 & 1119 & 1038 & -16.2 & Y \\
9 & 0.0325 & 147 & 4.9 & 7.8 & -7.1 & 7.9 & 64 & 236 & 1583 & 1348 & -15.4 & Y \\
10 & 0.038 & 172 & 3.2 & 6.8 & -6.5 & 7.8 & 49 & 195 & 1202 & 1283 & -15.9 & Y \\
11 & 0.0377 & 171 & 1.8 & 8.2 & -7.5 & 7.8 & 76 & 193 & 1103 & 1086 & -15.0 & Y \\
12 & 0.0225 & 101 & 1.4 & 6.4 & -6.9 & 7.9 &  & 251 & 1896 & 1460 & -15.6 & Y \\
13 &  & 143 &  & 7.6 & -8.4 &  &  &  &  &  &  & N \\
14 & 0.0327 & 148 & 1.9 & 7.3 & -7.4 & 7.9 & 36 & 289 & 2062 & 1296 & -15.8 & Y \\
15 &  & 187 & 2.4 & 7.9 & -7.7 &  &  &  &  &  &  & N \\
16 &  & 128 & 4.8 & 7.1 & -6.4 &  &  &  &  &  &  & N \\
17 &  & 185 &  & 7.8 & -7.6 &  &  &  &  &  &  & Y \\
18 & 0.0331 & 150 & 2.0 & 7.0 & -7.1 & 7.7 & 60 & 168 & 856 & 853 & -15.7 & Y \\
19 &  & 150 & 2.4 & 7.6 & -7.5 &  &  &  &  &  &  & Y \\
20 & 0.0203 & 91 & 4.3 & 7.0 & -7.4 & 8.1 &  & 296 & 1893 & 1284 & -16.0 & N \\
21 &  & 116 & 2.1 & 7.1 & -6.9 &  &  &  &  &  &  & Y \\
22 & 0.0227 & 102 & 5.4 & 8.3 & -6.6 &  &  & 237 &  &  & -14.7 & Y \\
23 &  & 130 & 3.3 & 7.8 & -8.1 &  &  &  &  &  &  & Y \\
24 &  & 126 & 6.4 & 6.4 & -7.0 &  &  &  &  &  &  & N \\
25 & 0.0224 & 101 & 1.6 & 6.2 & -6.9 & 7.9 &  & 255 & 1301 & 1279 &  & Y \\
26 &  & 177 & 2.1 & 6.7 & -8.0 &  &  &  &  &  &  & Y \\
27 & 0.0404 & 184 & 2.7 & 7.7 & -7.1 & 7.9 & 54 & 104 & 583 & 548 & -15.7 & Y \\
28 &  & 93 & 3.4 & 8.1 & -8.7 &  &  &  &  &  &  & Y \\
29 &  & 108 &  & 7.7 & -7.8 &  &  &  &  &  &  & Y \\
30 & 0.0328 & 148 & 2.2 & 7.4 & -7.5 & 7.8 & 50 & 129 & 638 & 743 & -15.2 & Y \\
31 & 0.0433 & 197 & 3.4 & 7.9 & -7.1 & 8.1 & 79 & 76 & 362 & 379 & -15.7 & N \\
32 & 0.0274 & 123 & 5.0 & 8.8 & -8.0 & 8.0 & 73 & 148 & 1005 & 792 & -15.1 & Y \\
33 & 0.0277 & 125 & 3.5 & 7.8 & -8.0 & 8.6 & 75 & 132 & 839 & 703 & -15.6 & N \\
34 & 0.0263 & 118 & 7.0 & 7.0 & -7.0 & 8.3 &  & 134 & 649 & 685 & -16.1 & N \\
35 &  & 95 &  & 7.4 & -8.0 &  &  &  &  &  &  & Y \\
36 & 0.0225 & 101 & 1.1 & 6.6 & -7.1 & 7.8 &  & 168 & 883 & 795 & -15.9 & Y \\
37 &  & 122 &  & 6.5 & -7.1 &  &  &  &  &  &  & N \\
38 &  & 140 & 2.9 & 7.7 & -6.3 &  &  &  &  &  &  & Y \\
39 &  & 155 &  & 6.9 & -6.4 &  &  &  &  &  &  & N \\
40 &  & 161 & 4.6 & 6.6 & -7.0 &  &  &  &  &  &  & N \\
41 &  & 162 & 5.4 & 7.2 & -7.2 &  &  &  &  &  &  & Y \\
42 &  & 18 & 1.3 & 7.1 & -8.7 &  &  &  &  &  &  & N \\
43 &  & 181 & 5.2 & 7.7 & -8.0 &  &  &  &  &  &  & N \\
44 &  & 85 & 2.7 & 6.8 & -7.0 &  &  &  &  &  &  & Y \\
45 &  & 162 & 4.5 & 6.7 & -6.5 &  &  &  &  &  &  & N \\
46 &  & 130 & 2.1 & 7.4 & -7.6 &  &  &  &  &  &  & N \\
47 &  & 124 & 1.3 & 7.3 & -7.0 &  &  &  &  &  &  & Y \\
48 &  & 107 & 4.0 & 6.9 & -7.1 &  &  &  &  &  &  & N \\
    \end{tabular}
    \begin{tablenotes}
      \small
      \item "MPA" refers to results provided by MPA-JHU, "HEC" to HECATE, "S" to spectral fitting, and "SED" to SED fitting.
      \item Distances are as reported in HECATE except for objects with spectroscopic observations, where distances are based on their redshift.
      \item Y=Yes and N=No if the BB is compact based on visual inspection. 
      \item Uncertainties are provided in the electronic form of the table.
    \end{tablenotes}
  \end{threeparttable}
    \label{tab:Blueberries_tab2}
\end{table*}

\subsection{Sizes and morphology}
\label{sec:sizes_morphology}

\begin{figure}
    \centering
    \includegraphics[width=\columnwidth]{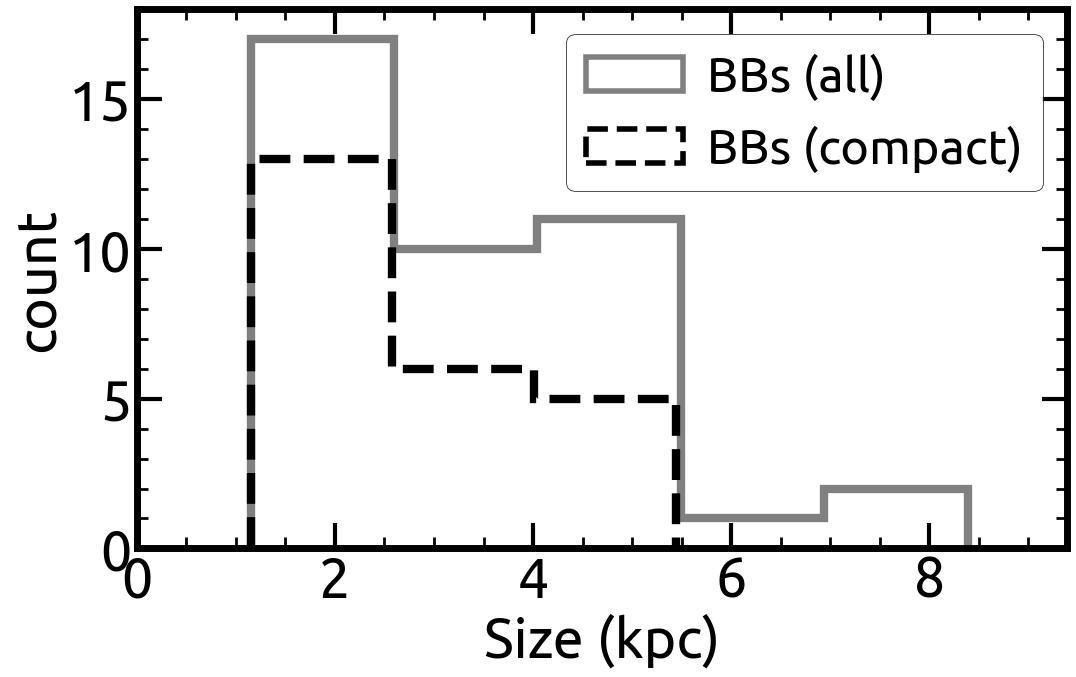}
    \caption{Distribution of sizes of the BBs found by this work.
    Sources that appear to have a compact morphology and no extended features are shown separately with a black dashed line, while the full sample is shown with a gray line.}
    \label{fig:dist_sizes}
\end{figure}

The size distribution of the selected BBs is shown in Figure \ref{fig:dist_sizes}.
The sizes were estimated based on the major axes reported in HECATE and the distances of the galaxies as reported in Table \ref{tab:Blueberries_tab2}.
The mode of their size distribution is $1.87^{+2.62}_{-0.53}$~kpc.
Visual inspection of this work's BBs (Figure \ref{fig:opt_images}) shows that the majority of BBs appear to be compact and isolated sources with a spherical morphology. 
However, some sources (BBs~1, 2, 3, 5, 6, 13, 15, 16, 20, 24, 31, 33, 34, 37, 39, 40, 42, 43, 45, 46, 48) show an extended morphology.
Some of these BBs may be blue star-forming regions that are part of a slightly larger galaxy, where a part of it has less SFR in comparison (e.g., BBs~3, 5, 6, 33, 34, 42).
For instance, this is indeed the case for BB~5, which has spectroscopic observations for the bluest ($\rm EW_{H\alpha} = 577$~\AA; northeast) and the less intensive ($\rm EW_{H\alpha} = 73$~\AA; southwest) regions of the source, both of which share the same redshift.
The optical and IR photometries correspond to the full extent of the galaxies. Thus, these sources fulfill the selection criteria even when accounting for their less intensive regions.
In some cases, nearby angular-projection sources may be background and thus unrelated galaxies (e.g., possibly BB~46).
Spectroscopic observations at different locations of the surface of these extended features would be required to identify whether these regions belong to these galaxies.

The mode and CIs of the distribution of distances for the compact BBs is $\left < D \right > = 138 \pm 38$~(Mpc), and for the ones with extended features, it is $\left < D \right > = 130^{+51}_{-23}$~(Mpc).
Therefore, we did not find a strong dependence on whether the BBs appear compact due to the combination of large distances and their small overall size. 
However, the two nearest sources show an extended morphology (BBs~3 and 42), Moreover, some recent studies have shown that some spatially resolved GP analogs show an irregular morphology \citep[e.g., Mrk 71 at 3.4~Mpc;][]{2017ApJ...845..165M}, while deep HST observations have also revealed extended structures in GPs \citep[e.g.,][]{2019ApJ...885...96J}.

The three largest reported BBs (6, 34, and 24) show extended features. 
Among the largest are also BBs~22 and 41 (size = 5.4~kpc) which appear to be compact sources.
However, these are exceptional sources, as the mode of the size distribution of compact BBs is $1.86^{+1.43}_{-0.52}$~kpc.
A unique source is BB~43, which appears to be a binary BB galaxy comprised of two compact sources that otherwise seem to be isolated from other nearby galaxies.
Unfortunately, there are no spectra available for BB~43, while its photometry accounts for both sources.

\subsection{Spectral energy distribution fitting}
\label{sec:SED_fitting}

To homogeneously estimate stellar masses and SFRs for all the reported BBs, we performed spectral energy distribution (SED) fitting, taking advantage of the multiwavelength photometric coverage of the sources.
In addition to the visible SDSS, Pan-STARRS, and IR WISE, the sources were cross-matched with the GALEX extended source catalog \citep{2005ApJ...619L...1M} in order to include UV photometry.
The galaxies BB~1, 3, 38, 42, 45, and 46 were not detected by GALEX in any of the AIS, GII, or MIS surveys.

The SED fitting was performed with the updated version \citep[][]{2022ApJ...927..192Y} of the widely used code \texttt{CIGALE} \citep{2005MNRAS.360.1413B,2009A&A...507.1793N,2019A&A...622A.103B}.
In particular, we used the \texttt{sfhdelayed}, \texttt{bc03}, \texttt{nebular}, \texttt{dustatt\_modified\_starburst}, \texttt{redshifting}, and \texttt{dale2014} modules covering a wide range of the parameter space.
We focused the parameter grid on low-metallicity and high sSFR galaxies, which is the case of BBs. 
For instance, the \texttt{sfhdelayed} module \texttt{f\_burst} parameter, which refers to the relative strength of the recent star-formation burst, was allowed up to 0.99 and the \texttt{age\_burst} was allowed down to 1~Myr.
We also allowed for a 10\% systematic uncertainty in the observed fluxes.
All sources were fitted with components only related to star formation.
Overall, the applied grid resulted in 5,225,472 models for each BB galaxy, and the best fit was chosen based on the minimum reduced $\chi^2$ ($\chi^{2}_{\nu}$).
All sources had good fits with the average $\langle \chi^{2}_{\nu} \rangle = 1.1$ and highest $\chi^{2}_{\nu}= 3.7$ for BB~20.
However, our SFR and $M_\star$ estimations were not based on the best-fit model but on their $\chi^{2}$-weighted probability density functions (PDFs).
Figure \ref{fig:CIGALE_fits} shows the SED plots of BB~20 and BB~30 as indicative of the SED fitting process.

Appendix \ref{sec:Appendix_B} provides the $\chi^{2}$-weighted PDFs of the SFR and $M_\star$ for each galaxy separately.
These PDFs cover all the models created by CIGALE in the fitting process, and thus, this approach allowed us to evaluate how much the SFR and $M_\star$ estimations vary for the different models and obtain a more robust estimation of the parameter quantities and uncertainties.
The quantity and error estimations provided by \cite{2022ApJ...927..192Y} correspond to the standard deviation of these PDFs.
However, because in our case most of the PDFs do not have a Gaussian form, we adopted and calculated the estimations based on the mode and 68\% CIs.

Since MPA-JHU also provides stellar masses and SFRs, we compared these estimations with our SED fitting for the 21 BBs of this work analyzed by MPA-JHU.
The logarithm of the ratio of the SED fitting and MPA-JHU SFRs is $\left< \rm log~(SFR_{SED}/SFR_{MPA}) \right> = 0.42^{+0.14}_{-0.16}$.
Similarly, compared to the HECATE SFRs, $\left< \rm log~(SFR_{SED}/SFR_{HECATE}) \right> = 0.64^{+0.11}_{-0.11}$.
Thus, the SFR estimations of the SED fitting tend to give larger SFRs.
The logarithmic ratio of the SED fitting and MPA-JHU stellar masses is $\left< \rm log~(M_{\star, SED}/M_{\star,MPA}) \right> = -0.22^{+0.24}_{-0.28}$ and $\left< \rm log~(M_{\star, SED}/M_{\star, HECATE}) \right> = -0.83^{+0.55}_{-0.09}$.
Thus, the stellar masses derived by the SED fitting tend to agree more with MPA-JHU and less with HECATE; however, there is considerable scatter.

The different methods presented here for the SFR and $M_\star$ estimations may suffer from the same or other biases.
Galaxies such as the BBs may not be very well represented by the \texttt{CIGALE} assumption of a standard \ion{H}{II} cloud with electron density $n_e = 100~{\rm cm}^-3$ and electron temperature $T_e =10^4$~K.
For instance, the nearest GP analog Mrk~71 shows $n_e = 235~{\rm cm}^-3$ and $T_e = 1.6 ~ 10^4$~K \citep{2017ApJ...845..165M}.
On the other hand, MPA-JHU SFRs were estimated through a combination of the emission-line measurements from the fiber spectra, with aperture corrections that were applied with the method of \cite{2005MNRAS.362...41G}, and the photometric measurements outside the fiber using the method of \cite{2007ApJS..173..267S}.
Only integrated photometry was used for sources with no adequate SFR estimation from the emission lines.
The SFRs and stellar masses in HECATEv2.0 are based on IR WISE photometry and combinations with optical photometry or estimations of extinction, depending on availability \citep{2023A&A...673A..16K}.
It is possible that in some cases MPA-JHU may have underestimated the SFR of the BBs since they did not include UV or mid-IR emission and may not have fully accounted for extinction.
Both are important given that BBs show strong UV emission. 
The IR-based estimations of the HECATE SFRs may underestimate the SFRs of high-sSFR and low-extinction galaxies such as the BBs for the same reason \citep[e.g.,][]{2023A&A...673A..16K}.

\subsection{Blueberry galaxies with respect to the main sequence of star-forming galaxies}
\label{sec:main_sequence}

\begin{figure}
    \centering
    \includegraphics[width=\columnwidth]{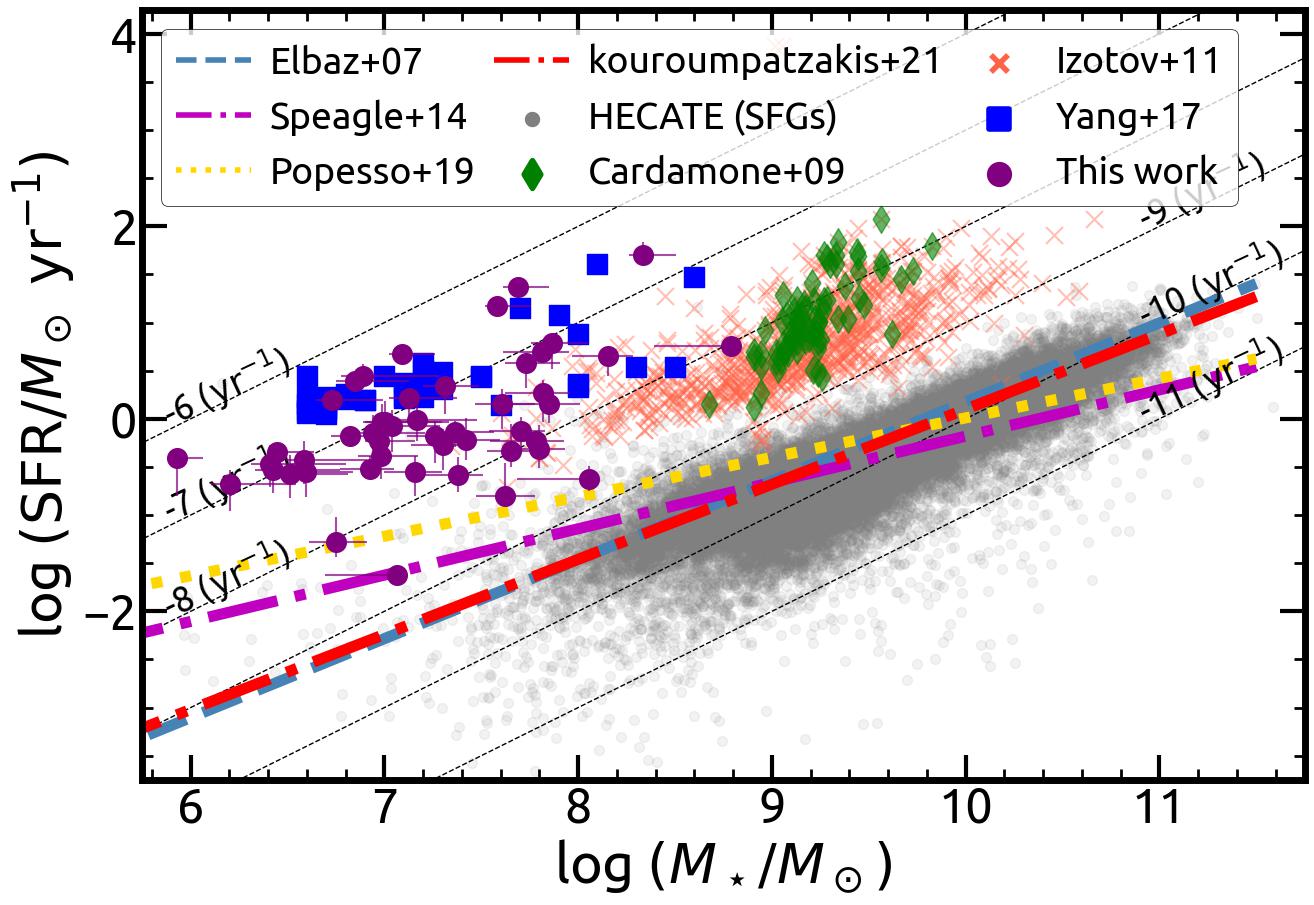}
    \caption{Star-formation rate as a function of the stellar mass of the BBs of this work (purple circles), the BBs of \protect\cite{2017ApJ...847...38Y} (blue squares), the GPs of \cite{2009MNRAS.399.1191C} (green rhombuses), the BDGs of \cite{2011ApJ...728..161I} (red X's), and the SFGs of HECATE (gray dots).
    The blue dashed, magenta dashed-dotted, yellow dotted, and red dashed-dotted lines respectively show the main sequence fits of
    \protect\cite{2007A&A...468...33E}, \protect\cite{2014ApJS..214...15S}, \protect\cite{2019MNRAS.483.3213P}, and \protect\cite{2021MNRAS.506.3079K}.
    The thin black dashed lines indicate some constant values of log~sSFR.}
    \label{fig:Main_Sequence}
\end{figure}

To place the proposed BBs in the context of their current star-forming conditions, we plotted their SFR as a function of $M_\star$, compared to the main sequence of SFGs.
Figure \ref{fig:Main_Sequence} shows the proposed BBs; the samples of \cite{2009MNRAS.399.1191C}, \cite{2011ApJ...728..161I}, and \cite{2017ApJ...847...38Y}; and all spectroscopically classified SFGs (\texttt{class\_sp = 0}) of the HECATE catalog.
The SFRs and stellar masses plotted in the figure refer to the results of the SED fitting for the BBs of this work (Section \ref{sec:SED_fitting}).
For the \cite{2017ApJ...847...38Y} BBs, SFRs, and stellar masses, they are as reported in that work.
For the HECATE catalog and the samples of \cite{2009MNRAS.399.1191C} and \cite{2011ApJ...728..161I}, the SFRs and stellar masses are as reported in HECATEv2.0 and the MPA-JHU catalogs, respectively.

All the aforementioned samples of dwarf galaxies show very high values of sSFR ($> 10^{-9}~{\rm yr^{-1}}$) and are placed above the main sequence compared to HECATE galaxies, which represent here the average population of SFGs in the near Universe.
Moreover, BBs tend to be even more extreme compared to the \cite{2009MNRAS.399.1191C} GPs and \cite{2011ApJ...728..161I} BDGs, with the majority being less massive and having $\rm sSFR > 10^{-8}~yr^{-1}$. 

\subsection{Visible colors of blueberry galaxies}
\label{sec:discuss_opt_properties}

\begin{figure}
    \centering
    \includegraphics[width=\columnwidth]{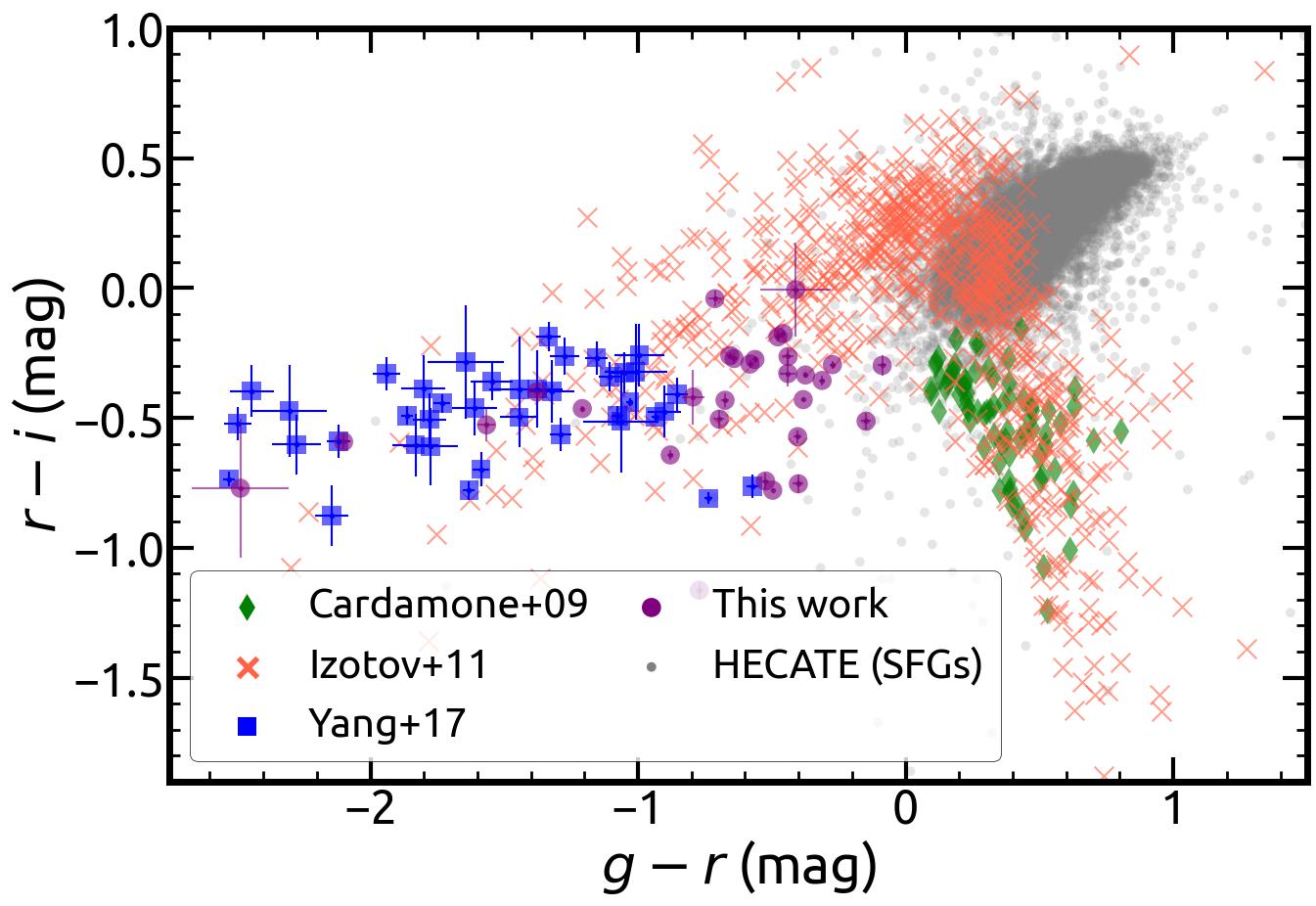}
    \caption{Color-color $r-i$ versus $g-r$ diagram with applied $k$-corrections.
    Gray dots, green rhombuses, red X's, blue squares, and purple circles respectively show sources of the HECATE SFGs, the \protect\cite{2009MNRAS.399.1191C}, the \protect\cite{2011ApJ...728..161I}, the \protect\cite{2017ApJ...847...38Y}, and this work's samples.
    Uncertainties are shown only for the BBs for clarity.}
    \label{fig:Opt_colors}
\end{figure}

Figure \ref{fig:Opt_colors} shows the rest-frame optical color-color $r-i$ versus $g-r$ diagram for this work's BBs, the \cite{2017ApJ...847...38Y} BBs, the \cite{2011ApJ...728..161I} BDGs, and the \cite{2009MNRAS.399.1191C} GPs.
The HECATE galaxies classified as star forming are also shown, representing the general population of SFGs in the nearby Universe.
The colors plotted for both BB samples and the HECATE SFGs are based on SDSS \texttt{cModelMag} photometry.
The colors of the BDGs and GPs are based on \texttt{ModelMag} photometry from SDSS DR12 \citep[][]{2015ApJS..219...12A}.
The photometry of all sources was $k$-corrected using the conversions of \cite{2010MNRAS.405.1409C} and \cite{2012MNRAS.419.1727C}.
Figure \ref{fig:Opt_colors} shows that some BBs are redder than $g-r = -0.4$, which is the selection threshold adopted by this work for SDSS photometry.
However, these sources were selected through Pan-STARRS photometry, which, as discussed in Section \ref{sec:criteria}, can be different compared to the \texttt{cModelMag} plotted here (see also Table \ref{tab:Blueberries_tab1}).
Colors reported in Table \ref{tab:Blueberries_tab1} refer to the observed (not the $k$-corrected) colors.

The majority of this work's BBs tend to be slightly redder in the $g-r$ color compared to \cite{2017ApJ...847...38Y} BBs but extend similarly in the $r-i$ color.
However, there are a few sources that are located at the blue-end of the $g-r$ color and extend to the \cite{2017ApJ...847...38Y} BBs.
Overall, this work's BBs tend to be located in the $g-r$ versus $r-i$ diagram between the \cite{2017ApJ...847...38Y} BBs and the \cite{2011ApJ...728..161I} BDGs and \cite{2009MNRAS.399.1191C} GPs.
This comparison also shows that the GPs and BDGs are on average intrinsically redder compared to BBs, although a small fraction of the large sample of \cite{2011ApJ...728..161I} BDGs share similar blue colors with the BBs.
Thus, the distinctive visible colors of BBs and GPs are not only different due to redshifting but are also intrinsically different.

\subsection{Spectroscopic properties of blueberry galaxies}
\label{sec:Spectroscopic_properties}

\begin{figure*}
    \centering
    \includegraphics[width=0.95\textwidth]{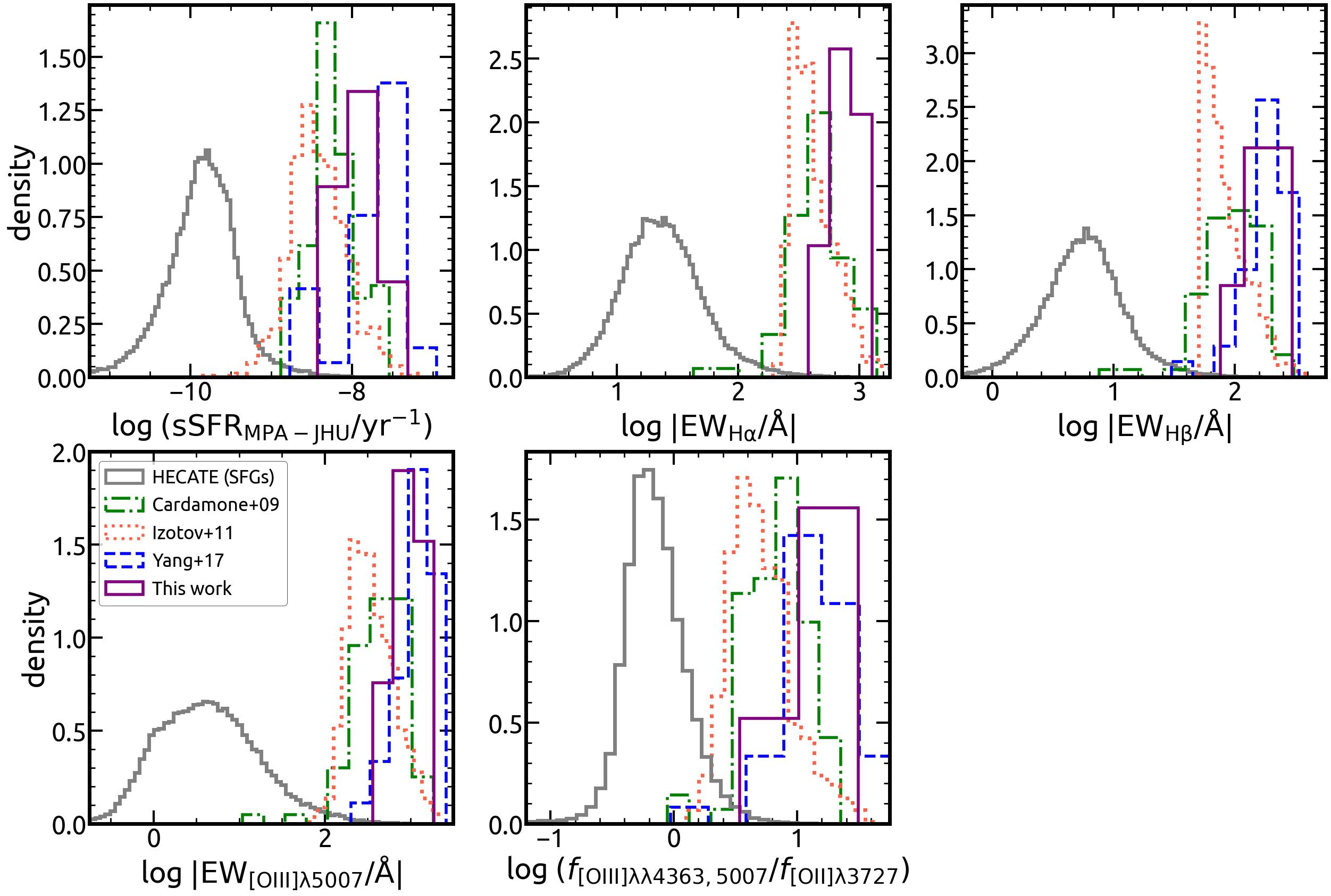}
    \caption{Histograms of (from left to right and top to bottom) the logarithm of the sSFRs; the EWs of $\rm H \alpha$, $\rm H \beta$, and $\rm [\ion{O}{III}] \lambda 5007$; and the flux ratios of $[\ion{O}{III}] \lambda \lambda 4363,5007 / [\ion{O}{II}] \lambda 3727 $ of the samples of \cite{2009MNRAS.399.1191C} (green dashed-dotted line), \cite{2011ApJ...728..161I} (red dotted line), \cite{2017ApJ...847...38Y} (blue dashed line), the SFGs of HECATE (gray line), and the spectroscopic subsample of this work's BBs (purple continuum line).}
    \label{fig:spec_results}
\end{figure*}

To examine their spectroscopic properties, the BBs were cross-matched with the MPA-JHU catalog, which provides robust results based on a homogeneous spectroscopic analysis.
However, this analysis could only be performed for the sources also covered by the MPA-JHU catalog, that is 21/48 galaxies.
The EWs of the $[\ion{O}{II}]\lambda 3726$, H$\beta$, $[\ion{O}{III}]\lambda 5007$, and H$\alpha$ emission lines and the metallicities for the spectroscopic subsample of the proposed BBs are reported in Table \ref{tab:Blueberries_tab2} and are based on the MPA-JHU catalog.
The distributions of the sSFRs and the EWs of the H$\beta$, $[\ion{O}{III}]\lambda 5007$, and H$\alpha$ emission lines for the GP and BB samples are presented in Figure \ref{fig:spec_results}.
The spectroscopically classified HECATE SFGs are also shown. 
Emission-line fluxes and EWs for the \cite{2009MNRAS.399.1191C}, \cite{2011ApJ...728..161I}, HECATE, and BBs of this work are as provided by the MPA-JHU analysis. 
Spectroscopic information of the \cite{2017ApJ...847...38Y} BBs is as reported in their work, but because many of these sources were observed with poor flat-fielding of the red part of the spectrum, their $\rm EWs_{H\alpha}$ were not properly estimated and are not shown in Figure \ref{fig:spec_results}.
Table \ref{tab:Spec_av_table} presents the modes and 68\% CIs of these distributions, summarizing the results shown in Figure \ref{fig:spec_results}.

The BB galaxies of this work are in agreement and show an sSFR distribution similar to that of the BBs of \cite{2017ApJ...847...38Y}. 
Both samples are on the right end of the sSFR distribution, and they show the highest star-forming activity among the presented samples. 
While the \cite{2009MNRAS.399.1191C} GPs and \cite{2011ApJ...728..161I} BDGs also show very high sSFRs ($\rm log ~ (sSFR/yr^{-1}) > -9$) compared to the main population of SFGs, BBs show even higher star-formation intensity.
An examination of the sample's EW distributions revealed they had similar behavior to those of the sSFRs.
This work's BBs show an EW distribution that is similar to the \cite{2017ApJ...847...38Y} BBs. 
Both samples of BBs have the highest EWs in any of the compared emission lines.

\begin{table*}
    \renewcommand{\arraystretch}{1.5}
    \centering
    \caption{Modes and 68\% CIs of the distributions of sSFRs; EWs of $\rm H \alpha$, $\rm H \beta$, and $\rm [\ion{O}{III}] \lambda 5007$; the flux ratio of $[\ion{O}{III}] \lambda \lambda 4363,5007 / [\ion{O}{II}] \lambda 3727 $; the metallicity; and the Balmer $E(B-V)$ of the samples of \cite{2009MNRAS.399.1191C}, \cite{2011ApJ...728..161I}, \cite{2017ApJ...847...38Y}, the SFGs of HECATE, and the spectroscopic subsample (21 sources) of this work's BBs.
    These quantities are also provided separately for the compact BBs (15 sources) and those with extended morphological features (six sources).
    These estimations are based on data provided by MPA-JHU.
    The last two rows present the modes and 68\% CIs of the distributions of the W1-W2 and W2-W3 colors based on ALLWISE data for all sources of the samples with $S/N>2$ in the WISE bands 1, 2, and 3.}
    \begin{threeparttable}
    \begin{tabular}{c|ccc|cccc}
        & BBs & BBs (ext.) & BBs (comp.) & BBs (Y) & GPs  & BDGs & SFGs\\
        \hline
         $\rm \left< log~({sSFR}/{yr^{-1})} \right>$ & $-7.87^{+0.18}_{-0.14}$ & $-8.13^{+0.0}_{-0.0}$ & $-7.87^{+0.12}_{-0.14}$ & $-7.5^{+0.03}_{-0.23}$ & $-8.32^{+0.14}_{-0.09}$ & $-8.55^{+0.19}_{-0.09}$ & $-9.8^{+0.15}_{-0.18}$\\
         $\rm \left< log~(EW_{H\alpha}/{\AA}) \right>$ & $2.84^{+0.05}_{-0.0}$ & $2.71^{+0.12}_{-0.0}$ & $2.95^{+0.08}_{-0.05}$ & & $2.67^{+0.08}_{-0.08}$ & $2.45^{+0.11}_{-0.02}$ & $1.39^{+0.13}_{-0.15}$\\
         $\rm \left< log~(EW_{H\beta}/{\AA}) \right>$ &  $2.18^{+0.13}_{-0.06}$ & $2.03^{+0.09}_{-0.0}$ &  $2.24^{+0.07}_{-0.01}$ & $2.26^{+0.08}_{-0.07}$ & $2.04^{+0.1}_{-0.12}$ & $1.73^{+0.1}_{-0.01}$ & $0.77^{+0.12}_{-0.14}$\\
         $\rm \left< log~(EW_{[\ion{O}{III}] \lambda 5007}/{\AA}) \right>$ & $2.92^{+0.08}_{-0.11}$ & $2.74^{+0.11}_{-0.0}$ & $3.04^{+0.03}_{-0.09}$ & $3.09^{+0.09}_{-0.07}$ & $2.65^{+0.13}_{-0.13}$ & $2.34^{+0.17}_{-0.05}$ & $0.6^{+0.27}_{-0.26}$\\
          $\left< {\rm log}~ (f_{[\ion{O}{III}] \lambda \lambda 4363,5007} / f_{[\ion{O}{II}] \lambda 3727}) \right>$ & $1.26^{+0.0}_{-0.1}$ & $1.28^{+0.0}_{-0.19}$ & $1.27^{+0.21}_{-0.08}$ & $1.05^{+0.16}_{-0.07}$ & $0.92^{+0.07}_{-0.15}$ & $0.57^{+0.14}_{-0.06}$ & $-0.21^{+0.11}_{-0.09}$\\
          $\rm \left< 12 + log(O/H) \right >$ &  $7.88^{+0.22}_{-0.09}$ & $7.86^{+0.0}_{-0.19}$ &  $8.12^{+0.0}_{-0.34}$  & $7.72^{+0.11}_{-0.12}$ & $8.82^{+0.05}_{-0.22}$ & $8.58^{+0.08}_{-0.17}$ & $9.0^{+0.05}_{-0.17}$\\
          $\left< E(B-V) \right >$~(mag) & $0.07^{+0.01}_{-0.01}$ & $0.03^{+0.0}_{-0.0}$ &  $0.03^{+0.03}_{-0.0}$ & & $0.07^{+0.03}_{-0.01}$ & $0.09^{+0.04}_{-0.02}$ &  $0.01^{+0.12}_{-0.01}$\\
          \hline
          $\rm \left< W1-W2 \right >$~(mag) & $1.24^{+0.08}_{-0.35}$ & $0.46^{+0.69}_{-0.08}$ & $0.75^{+0.49}_{-0.04}$ & $0.21^{+0.44}_{-0.22}$ & $0.59^{+0.36}_{-0.23}$ & $0.42^{+0.19}_{-0.08}$ & $0.1^{+0.04}_{-0.04}$\\
          $\rm \left< W2-W3 \right >$~(mag) & $4.7^{+0.16}_{-0.18}$ & $4.53^{+0.11}_{-0.08}$ & $4.69^{+0.16}_{-0.32}$ & $4.78^{+0.24}_{-0.38}$ & $4.12^{+0.1}_{-0.16}$ & $4.31^{+0.19}_{-0.19}$ & $3.37^{+0.17}_{-0.25}$\\
    \end{tabular}
        \begin{tablenotes}
      \small
      \item The modes and CIs of the sSFRs of this work's BBs mentioned in this table refer to the MPA-JHU estimations regarding the 21 spectroscopic sources. 
      The modes and CIs of the sSFRs of all 48 BBs based on the SED fitting is  $\rm \left< log~{sSFR}/{yr^{-1}} \right> = -7.03^{+0.12}_{-0.34}$.
      \end{tablenotes}
  \end{threeparttable}
    \label{tab:Spec_av_table}
\end{table*}

The distributions of log~$(f_{[\ion{O}{III}] \lambda \lambda 4363,5007} / f_{[\ion{O}{II}] \lambda 3727})$ for the different samples are shown in Figure \ref{fig:spec_results}. 
The ratio of the different oxygen ions is a good tracer of the interstellar medium's (ISM) ionization degree.
The distributions of GPs and BDGs overlap to some extent with the BBs, but the modes of their distributions have lower ratios. Thus, the BBs show the most ionized ISM among the compared samples.

\begin{figure}
    \centering
    \includegraphics[width=\columnwidth]{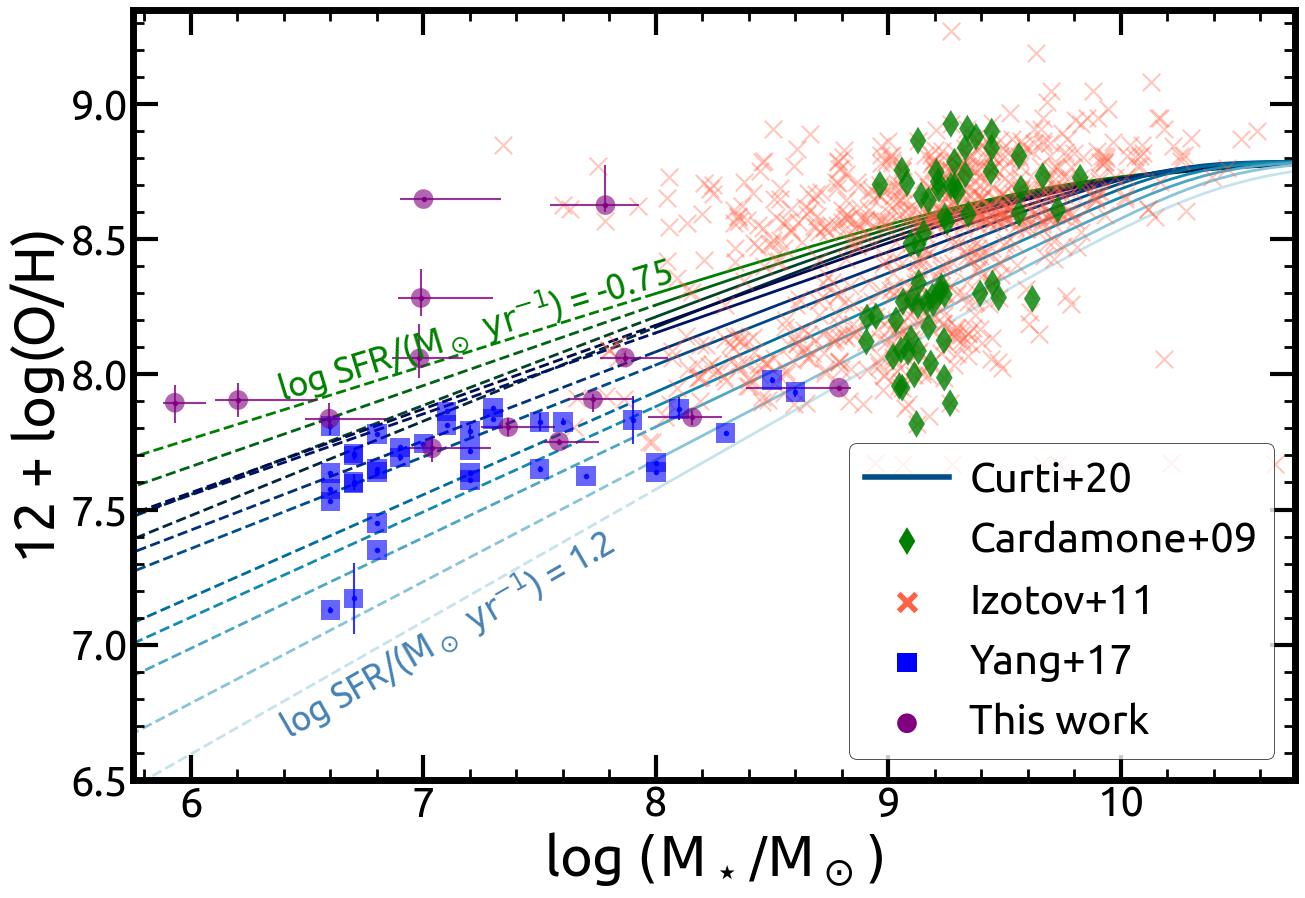}
    \caption{Mass--metallicity relation of the samples of \cite{2009MNRAS.399.1191C} GPs (green rhombuses), \cite{2011ApJ...728..161I} BDGs (red X's), \cite{2017ApJ...847...38Y} BBs (blue squares), and the spectroscopic subsample of this work's BBs (purple circles).
    Uncertainties are not shown for the first two samples for clarity.
    Colored curves show the mass-SFR-metallicity relation from Equation 2 of \protect\cite{2020MNRAS.491..944C} for a range of $\rm log~(SFR/M_\odot~yr^{-1}) = 1.2$ (green) to $\rm log~(SFR/M_\odot~yr^{-1}) = -0.75$ (blue).}
    \label{fig:mass_metallicity}
\end{figure}

The mass--metallicity relation of the BBs, GPs, and BDGs is presented in Figure \ref{fig:mass_metallicity}.
Metallicities and stellar masses of all samples, except the \cite{2017ApJ...847...38Y} BBs, are as reported by the MPA-JHU analysis, where the metallicity estimation was based on the \protect\cite{10.1046/j.1365-8711.2001.04260.x} models.
Stellar masses and metallicities of the \cite{2017ApJ...847...38Y} BBs are as reported in their work.
Most BBs show extremely low metallicities, with $\rm 12+log(O/H) < 8$.
The mode of the metallicity distribution of the \cite{2017ApJ...847...38Y} BBs is $\rm \langle 12+log(O/H) \rangle = 7.72^{+0.11}_{-0.12}$, and for this work's BBs, it is $\rm \langle 12+log(O/H) \rangle = 7.88^{+0.22}_{-0.09}$.
For reference, we also plot in Figure \ref{fig:mass_metallicity} the mass--SFR--metallicity relation by \cite{2020MNRAS.491..944C} provided in their Equation 2.
Estimations below $M_\star < 10^{8}~(M_\odot)$ are an extrapolation since the \protect\cite{2020MNRAS.491..944C} relation was a product of an analysis with more massive galaxies compared to BBs.

The highly star-forming and low-mass BBs tend to have metallicities close to what is expected from the relation accounting for their stellar mass and SFR.
For instance, the mode of the SFR distribution of the plotted BBs of this work is $\rm \langle log~(SFR/M_\odot~yr^{-1}) \rangle = -0.16^{+0.61}_{-0.87}$, while the minimum and maximum SFRs are $\rm log~(SFR/M_\odot~yr^{-1}) = -1.62$,  $\rm log~(SFR/M_\odot~yr^{-1}) = 1.7$, respectively. 
The upper line drawn from the \cite{2020MNRAS.491..944C} relation refers to $\rm log~(SFR/M_\odot~yr^{-1}) = -0.75$. Thus, the majority of the BBs are below the lines that correspond to their actual SFRs.
However, the comparison shows significant scatter.
Compared to BBs, GP and most BDG galaxies are more massive and show, on average, higher metallicities.

Two outliers also show a higher metallicity than the rest of the BBs (BBs 2 and 33).
Both of these sources show extended features rather than the typical compact morphology of the majority of the BBs, and their spectroscopic analysis refers to these regions.  
This suggests that these extended features are part of the same galaxy, which probably hosts older SPs and has an ISM with a higher metallicity.

\begin{figure}
    \centering
    \includegraphics[width=\columnwidth]{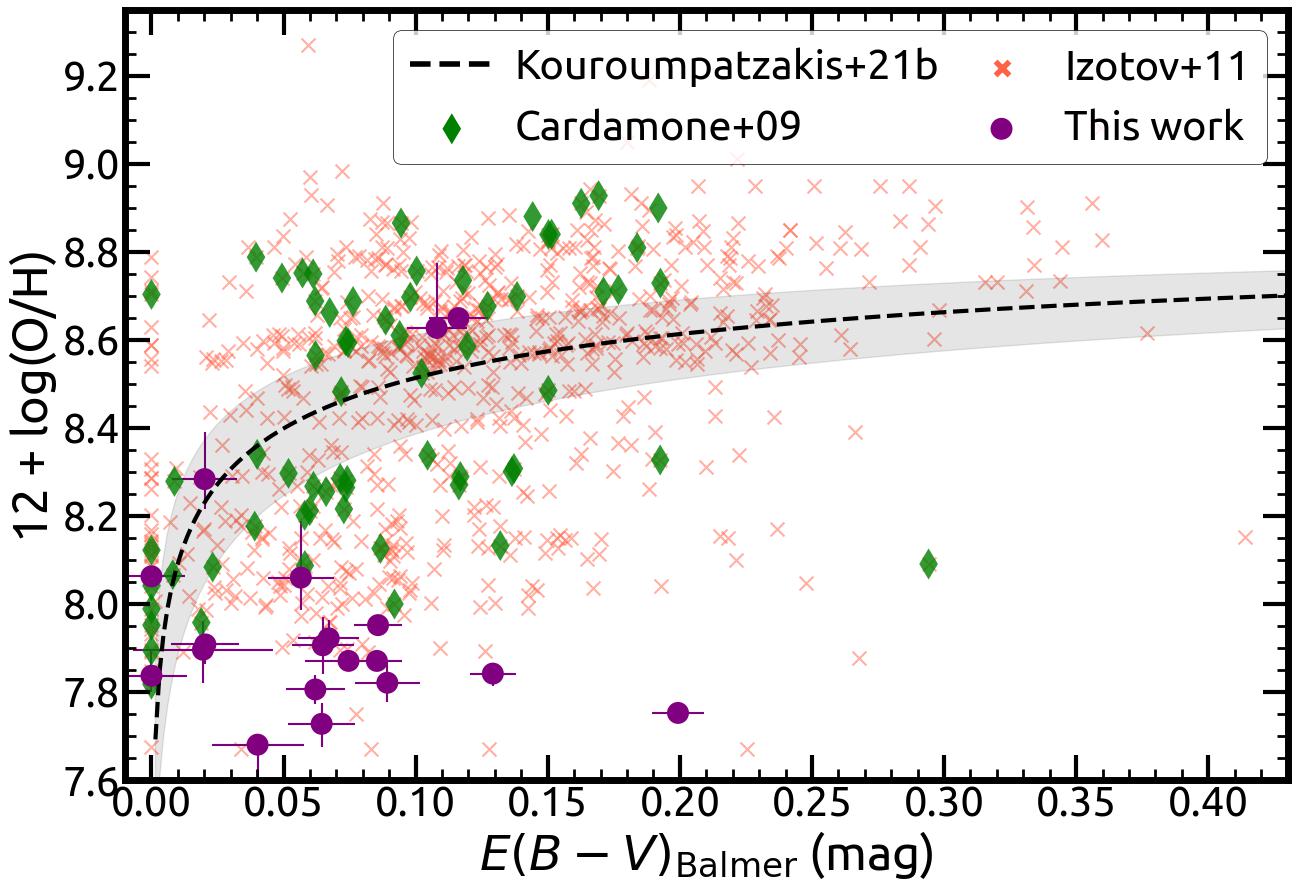}
    \caption{Metallicity as a function of extinction for the spectroscopic subsample of the proposed BBs (purple error bars). 
    The samples of \protect\cite{2009MNRAS.399.1191C} and \protect\cite{2011ApJ...728..161I} are shown with green rhombuses and red X's, respectively.
    Uncertainties are not shown for the latter for clarity.
    The black dashed line shows the metallicity-extinction relation of SFGs by \cite{2021MNRAS.506.3079K}.}
    \label{fig:metall_extinction}
\end{figure}

The comparison between the metallicity and the extinction of the aforementioned samples is shown in Figure \ref{fig:metall_extinction}.
The extinction in the figure is based on the Balmer ratio and the emission-line fluxes measured by MPA-JHU for the spectroscopic subsample of the proposed BBs. 
The $E(B-V)$ was calculated by adopting Case B recombination with $F_{\rm H\alpha}/F_{\rm H\beta} = 2.86$ \citep[][]{1989agna.book.....O} and the reddening law of \cite{2000ApJ...533..682C}.
The $E(B-V)$ of sources that were found to have a Balmer ratio $ f_{\rm H\alpha}/ f_{\rm H\beta} < 2.86$ was set to zero.
The respective extinction--metallicity relation by \cite{2021MNRAS.506.3079K}, which was based on a representative sample of all types of SFGs, is also drawn in the figure.
Figure \ref{fig:metall_extinction} does not show the \cite{2017ApJ...847...38Y} sample because for some galaxies, there were problems with the red part of their spectrum.
While most \cite{2009MNRAS.399.1191C} GPs and \cite{2011ApJ...728..161I} BDGs tend to follow the relation, given the large scatter, this is not the case for the majority of BBs.
Based on their metallicity, most of this work's BBs are below the relation and show stronger extinction than expected.

Table \ref{tab:Spec_av_table} also provides the modes and 68\% CIs of the distributions of the quantities discussed above, separately for the morphologically compact BBs and those showing extended features.
On average, compact BBs tend to form stars more actively and show slightly higher sSFRs and EWs.
On the other hand, compact BBs on average seem to have higher metallicities, although their distributions overlap, and the BB with the highest metallicity (BB~2; 12 + log(O/H) = 8.6) shows extended features. 
Out of the 21 spectroscopically observed BBs, only six sources are reported as extended and 15 as compact. Thus, these results should be taken with caution due to the small and uneven size of these subsamples.

The high ionization potential of $\rm \ion{He}{II}~\lambda 4686$ (54.42~eV) is considered in many cases an indication of AGN activity.
However, when an AGN is not responsible, such high ionization levels can only be achieved by the presence of very massive stars and recent bursts of star formation, which generate SPs with ages less than 5~Myr \citep[e.g.,][]{1997ApJS..108..339B}.
The $\rm \ion{He}{II}~\lambda 4686$ emission-line fluxes were measured by \cite{2012MNRAS.421.1043S} for the spectra of SDSS Data Release 7.
In total, eight BBs of this work were reported by \cite{2012MNRAS.421.1043S} as sources with detected $\rm \ion{He}{II}~\lambda 4686$ emission with S/N$>5.5$. 

To homogeneously estimate the $\rm \ion{He}{II}~\lambda 4686$ fluxes for all the BBs with spectroscopic observations, we fit their SDSS spectra with the SP fitting code \texttt{FADO} \citep{2017A&A...603A..63G}.
The \texttt{FADO} fits were based on the \cite{2003MNRAS.344.1000B} simple SP template library with metallicities $0.004 < Z < 0.05$ and $\rm 1 ~ Myr < SP_{age} < 15 ~ Gyr$.
All fits used the spectral range $3730 <$~\AA\ $<9000$, while we adopted the recommended resolution for SDSS spectra of 2.3~\AA\ and recessional velocity margins of 500~$\rm km~s^{-1}$.
The fluxes and EWs of all the emission lines estimated by \texttt{FADO} are provided in the online version of Table \ref{tab:Blueberries_tab2}.

A comparison between the flux estimations of our analysis and those of MPA-JHU showed excellent agreement, with $ \left < {\rm log} ~ (f_{\rm H\alpha-FADO}/f_{\rm H\alpha-MPA}) \right > = -0.15^{+0.2}_{-0.0}$,  $\left < {\rm log} ~ (f_{\rm H\beta-FADO}/f_{\rm H\beta-MPA})  \right > = 0.1^{+0.01}_{-0.03}$, and $\left < {\rm log} ~(f_{\rm [\ion{O}{III}]-FADO}/f_{\rm [\ion{O}{III}]-MPA}) \right > = -0.08^{+0.14}_{-0.0}$.
The comparison of the EW estimations is $ \left < \rm log ~ (EW_{\rm H\alpha-FADO}/EW_{\rm H\alpha-MPA})  \right > =  0.11^{+0.01}_{-0.02}$,  $\left < \rm log~ (EW_{\rm H\beta-FADO}/EW_{\rm H\beta-MPA}) \right > = 0.16^{+0.0}_{-0.01}$, and $\left < \rm log ~ (EW_{\rm [\ion{O}{III}]-FADO}/EW_{\rm [\ion{O}{III}]-MPA}) \right > = 0.03^{+0.01}_{-0.0}$.
Thus, FADO tends to yield slightly larger EWs on average, mainly for the Balmer lines.
Out of the 21 sources with spectroscopic observations, 15 were found to have $\rm \ion{He}{II}~\lambda 4686$ emission with $S/N>2$, showing that the BBs are indeed dominated by extremely young SPs.
In particular, these are BBs 2, 7, 9, 10, 11, 12, 14, 18, 22, 27, 30, 31, 32, 33, and 36.

\subsection{Infrared colors of blueberry galaxies}
\label{sec:discuss_IR_colors}

\begin{figure*}
    \centering
    \includegraphics[width=0.9\textwidth]{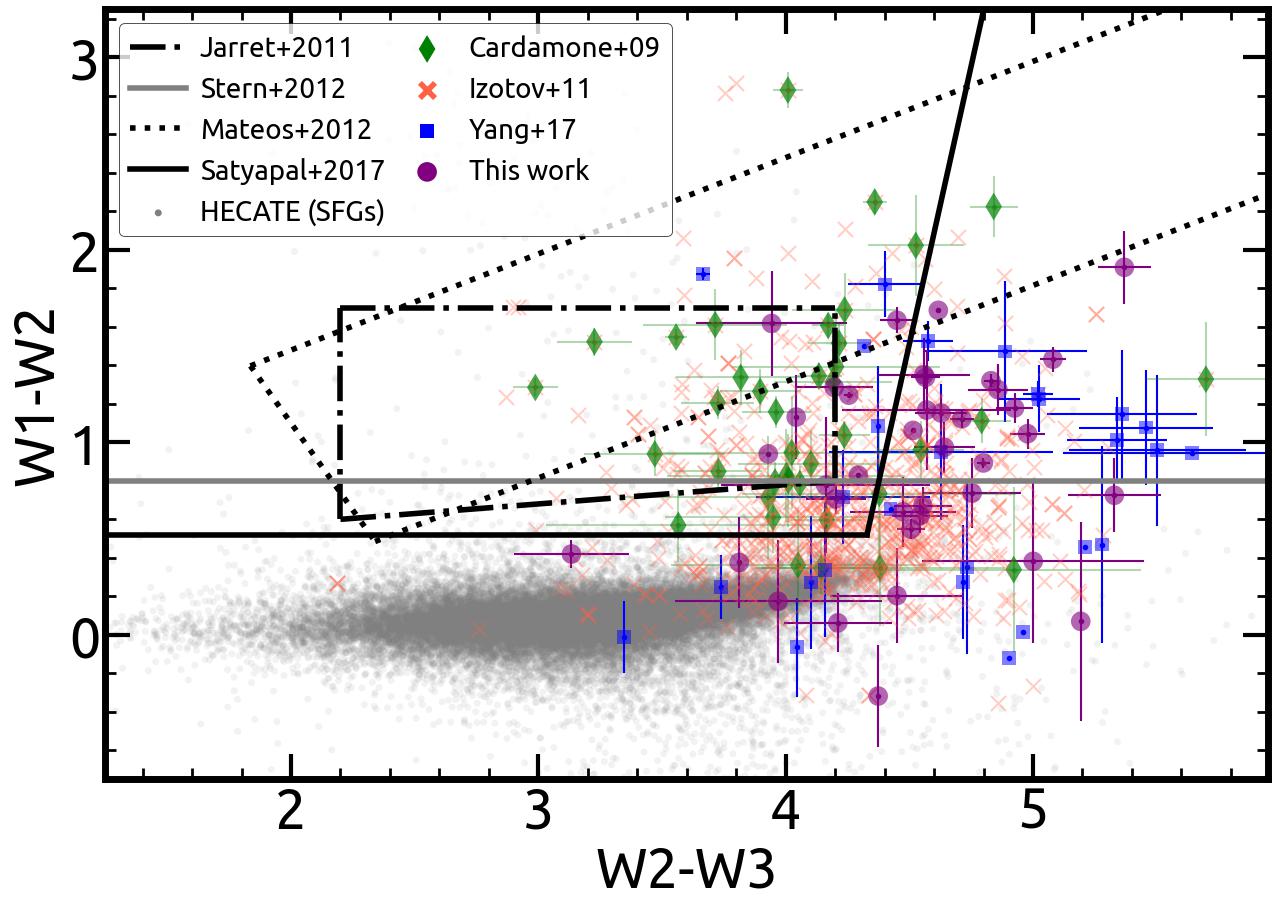}
    \caption{Color-color $\rm W1-W2$ versus $\rm W2-W3$ diagram for the samples of \protect\cite{2009MNRAS.399.1191C} (green rhombuses), \protect\cite{2011ApJ...728..161I} (red X's), \protect\cite{2017ApJ...847...38Y} (blue squares), this work's BBs (purple circles), and the SFGs of the HECATE catalog (gray dots).
    Uncertainties for the HECATE and \protect\cite{2011ApJ...728..161I} galaxies are not shown for greater clarity; however, all plotted sources have $S/N > 2$ in the used WISE bands 1, 2, and 3. 
    The black dashed-dotted, gray continuous, black dotted, and black continuous lines respectively show the \protect\cite{2011ApJ...735..112J}, \protect\cite{2012ApJ...753...30S}, \protect\cite{2012MNRAS.426.3271M}, and \protect\cite{2018ApJ...858...38S} criteria defined to identify AGNs and quasar sources.}
    \label{fig:IR_colors}
\end{figure*}

The WISE IR colors of the aforementioned samples are shown in Figure \ref{fig:IR_colors}.
The comparison shown in the figure involves 38 BBs of this work.
Also included are the 39, 719, and 27 galaxies respectively of the \cite{2009MNRAS.399.1191C}, \cite{2011ApJ...728..161I}, and \cite{2017ApJ...847...38Y} samples that were detected with an $S/N>2$ in WISE bands 1, 2, and 3 in the ALLWISE catalog \citep{2010AJ....140.1868W,2011ApJ...731...53M}.
We performed cross-matching with ALLWISE with a search radius of $5''$. 
Additionally, all HECATE galaxies detected with an $S/N>2$ in the same WISE bands and classified as star forming by MPA-JHU are also shown, representing the main population of SFGs.

Most BDGs, GPs, and BBs deviate significantly compared to the main population of SFGs, having significantly redder colors both in $\rm W1-W2$ and $\rm W2-W3$.
The BBs of this work have $\rm W1-W2 > -0.31 ~ (mag)$ and $\rm W2-W3 > 3.14 ~ (mag)$.
Similarly, the \cite{2017ApJ...847...38Y} BBs were found to have $\rm W1-W2 > 0.34 ~ (mag)$ and $\rm W2-W3 > 3.66 ~ (mag)$.
The majority of all the BBs have $\rm W1-W2 > 0.5 ~ (mag)$ and $\rm W2-W3 > 4 ~ (mag)$.
Compared to other dwarf SFGs, BBs show the reddest IR colors and are as red as the much dustier and more massive (ultra) luminous IR galaxies \citep[U-LIRGS;][]{2011ApJ...735..112J}.
A comparison between the BBs with extended photometry (17 sources with S/N$>2$) and the compact BBs (17 sources with S/N$>2$) showed that the latter have slightly redder IR colors, indicating more intense activity (Table \ref{tab:Spec_av_table}).
However, their IR color distributions overlap significantly.

The BDGs of the larger \cite{2011ApJ...728..161I} sample are spread over a broad area of the diagram, with some reaching an extremely red IR color (e.g., $\rm W1-W2 \simeq 2.8$), but most are between the BBs and the typical SFGs.
The \cite{2009MNRAS.399.1191C} GPs have $\rm W1-W2$ colors similar to BBs; however, they show relatively bluer $\rm W2-W3$ colors.

\begin{figure}
    \centering
    \includegraphics[width=\columnwidth]{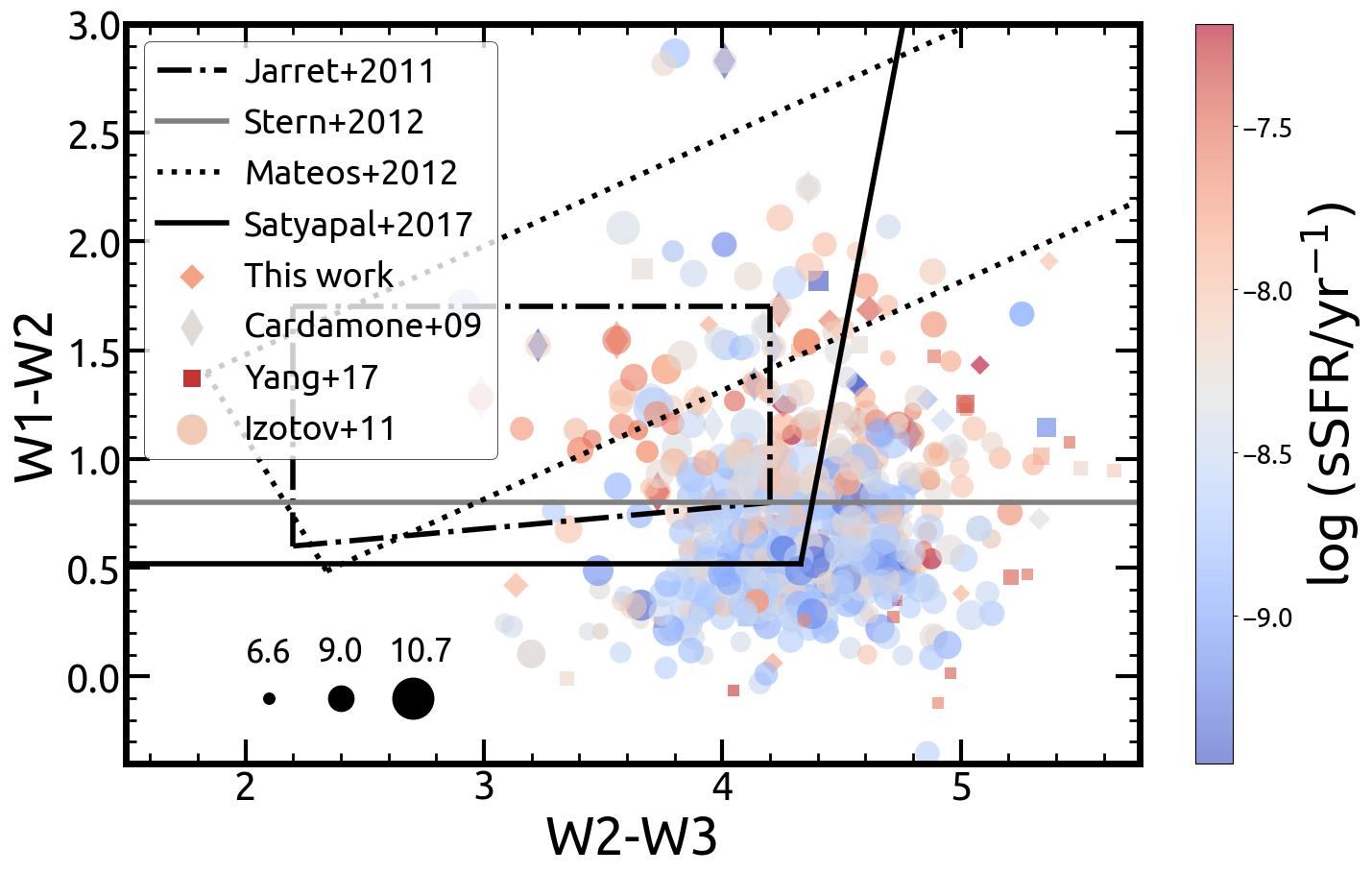}
    \caption{WISE $W1-W2$ versus $W2-W3$ diagram of BBs, GPs, and BDGs.
    All points are colored based on the galaxies' sSFRs.
    The samples of \protect\cite{2009MNRAS.399.1191C} GPs, \protect\cite{2011ApJ...728..161I} BDGs, \protect\cite{2017ApJ...847...38Y}, and this work's BBs are shown with rhombus, circle, square, and diamond markers, respectively.
    The size of the markers is based on the power of the logarithm of the stellar mass of each galaxy and is indicated in the bottom-left corner of the figure.
    The black dashed-dotted, gray continuous, black dotted, and black continuous lines respectively show the \protect\cite{2011ApJ...735..112J}, \protect\cite{2012ApJ...753...30S}, \protect\cite{2012MNRAS.426.3271M}, and \protect\cite{2018ApJ...858...38S} wedges used to identify AGNs and quasars.}
    \label{fig:IR_colors_sSFR}
\end{figure}

The dependence of the IR colors of BBs and BDGs on their star-forming intensity is examined in Figure \ref{fig:IR_colors_sSFR}. 
This figure presents a comparison that involves 28 galaxies of this work and 19, 576, and 27 galaxies respectively of the  \cite{2009MNRAS.399.1191C}, \cite{2011ApJ...728..161I}, and \cite{2017ApJ...847...38Y} samples with an $S/N>5$ in the WISE bands 1, 2, and 3.
Here, the sSFR is as reported in MPA-JHU for the \cite{2009MNRAS.399.1191C} and \cite{2011ApJ...728..161I} samples, while for the BBs of this work, the sSFR estimations come from the SED fitting (Section \ref{sec:SED_fitting}).
The sSFRs of the \cite{2017ApJ...847...38Y} BBs are as reported in that work.

As the color coding reveals, BDGs with a higher sSFR tend to have redder IR colors.
The population of BDGs shows a stronger color gradient with the $W1-W2$ color than with the $W2-W3$ color.
Most of the BDGs with $W1-W2 \gtrsim 1$ are galaxies with increased star-forming intensity (log~$\rm sSFR \gtrsim -8 ~ yr^{-1}$). 
However, there are a few high-sSFR but low-mass BBs that have $W1-W2 \simeq 0$.

\section{Discussion}
\label{sec:Discussion}

\subsection{Classification of the blueberry galaxies}
\label{sec:classification}

The optical emission-line diagram \citep[BPT;][]{1981PASP...93....5B,1987ApJS...63..295V,2001ApJ...556..121K,2003MNRAS.346.1055K} analysis of this sample classified all the proposed BBs as purely SFGs.
Similarly, the 4D classification scheme of \cite{2019MNRAS.485.1085S}, which simultaneously takes into account all four emission-line ratios of the traditional BPT diagram classified as star forming with $100\%$ probability all 21 spectroscopically classifiable BBs.

Additionally, to characterize the activity of all the sources of this sample, we applied the scheme proposed by \cite{2023A&A...679A..76D}, which is based on their photometric IR $\rm W1-W2$, $\rm W2-W3$ and the visible $g-r$ colors.
This method also returns the probability of each source belonging to any of the classes.
Again, based on this independent method, all 48 BBs were classified as SFGs.
However, only one galaxy, BB~5, was classified as an SFG with $p_{\rm SFG} > 75\%$, which is considered as secure estimation, while the rest have $57\% < p_{\rm SFG} < 75\%$.
Galaxy BB~31 (PGC~3108006) has the highest probability of being classified as an AGN, with $p_{\rm AGN}=26\%$, but it is still considered an SFG with a much higher probability ($p_{\rm SF}=57\%$).

Although most of the GPs and BBs are above the \cite{2012ApJ...753...30S} $\rm W1-W2$ color limit, all of these galaxies were classified as purely star forming based on the BPT diagram and the IR diagnostics of \cite{2023A&A...679A..76D}.
Additionally, a rather small fraction of the \cite{2011ApJ...728..161I} BDGs and a larger fraction of the \cite{2009MNRAS.399.1191C} GPs were placed inside the \cite{2011ApJ...735..112J} and \cite{2018ApJ...858...38S} wedges of AGN galaxies.
The criteria of \cite{2018ApJ...858...38S} were particularly defined to separate the IR colors of BDGs and AGNs.
To do so, they modeled with \texttt{Starburst99} BDGs with a continuous starburst at 5~Myrs and a very strong radiation field, and then they modeled the nebular and dust emission with \texttt{Cloudy} \citep{2013RMxAA..49..137F}.
However, a small fraction of IR-detected \cite{2011ApJ...728..161I} BDGs and the majority of \cite{2009MNRAS.399.1191C} GPs still fulfilled these color criteria, possibly due to complex SFHs and the presence of both old and young SPs, whose combined emission can soften the $\rm W2-W3$ color.

\cite{2023ApJ...945..157H} studied the IR variability of the BPT star-forming GPs falling within the \cite{2011ApJ...735..112J} criteria.
Two sources showed variability in their IR emission, which is a strong indication of the presence of an AGN.
Out of all BBs studied here, only BB~2 (PGC~5094058) falls within the \cite{2011ApJ...735..112J} wedge, close to its upper limit.
Overall, the majority of GPs and BBs are above the \cite{2012ApJ...753...30S} $W1-W2$ limit, but most \cite{2011ApJ...728..161I} BDGs are below it. 
Only a fraction of the BDGs, GPs, and BBs enter the \cite{2012MNRAS.426.3271M} wedge.
Moreover, while the \cite{2011ApJ...735..112J} and \cite{2018ApJ...858...38S} color criteria work well for the majority of BBs, they tend to misclassify GP galaxies.

\begin{figure}
    \centering
    \includegraphics[width=\columnwidth]{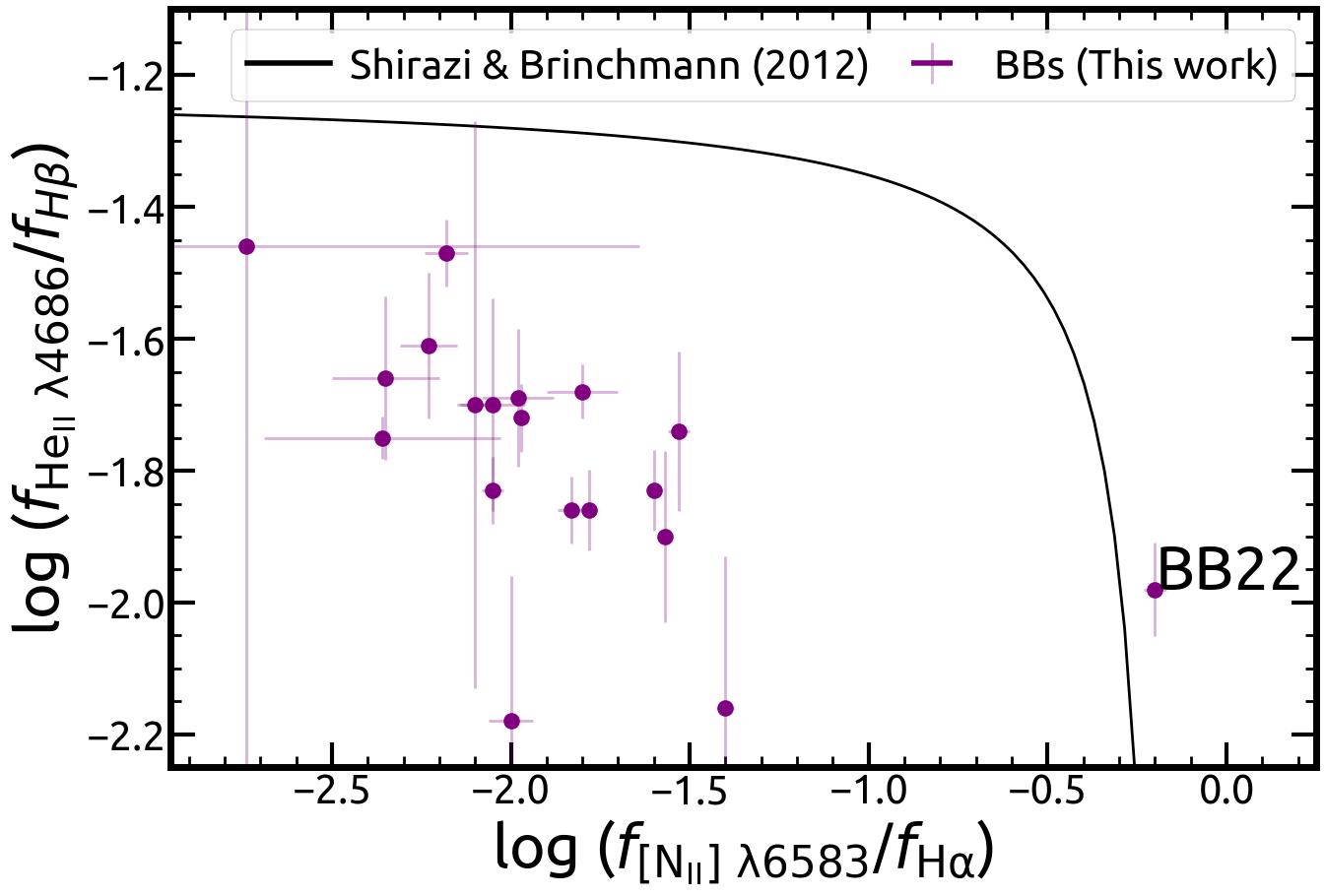}
    \caption{Emission-line of $\rm \ion{He}{II}~\lambda 4686 / H\beta$ against $\rm [\ion{N}{II}]~\lambda 6583/H\alpha$ flux ratios of the spectroscopically observed BBs of this work (purple error bars).
    The black line shows the \cite{2012MNRAS.421.1043S} relation.}
    \label{fig:HeII_plot}
\end{figure}

The galaxies BB~22 and BB~7 were reported as AGN galaxies by \cite{2012MNRAS.421.1043S}.
These sources were most likely classified as AGNs because they show slightly broader emission lines.
However, visual inspection of their spectra shows that both permitted and forbidden emission lines show broad features, which is a characteristic of outflows related to star-forming activity (e.g., supernovae, stellar winds, superbubbles) rather than an AGN.
Using our emission-line flux measurements (Section \ref{sec:Spectroscopic_properties}), we plot in Figure \ref{fig:HeII_plot} the $\rm \ion{He}{II} / H\beta$--$\rm [\ion{N}{II}]/H\alpha$ diagram for the BBs of this work.
Notably, BB~22 is the only source located in the AGN region of the diagram, while the rest of BBs are clearly below the \cite{2012MNRAS.421.1043S} relation.

In order to further examine the classification of the proposed BBs, we performed an archival search of their radio footprints.
While a majority of the sources were not observed by LOFAR DR2 \citep[][]{FIRST,LoTSS-DR2}, which covers part (27\%) of the northern sky in low 120--168 MHz frequencies, it successfully detected BBs~7, 8, 9, 21, 27, and 33, but BB~22 was not observed.
Among all BBs identified by this work, BB~22 is the only source with detectable radio emission in the radio surveys of the NRAO VLA Sky Survey \cite[NVSS;][]{NVSS}, Very Large Array Sky Survey \citep[VLASS;][]{VLASS}, and the Rapid ASKAP Continuum Survey \citep[RACS;][]{RACS-low} that operate at higher radio frequencies.
All other BBs were not detected or were not observed by the aforementioned radio surveys even though some should have been detected based on the general SFR--radio relation.
However, it has been shown that low-mass galaxies do not closely follow the general SFR-radio relations and are usually underluminous \citep[e.g.,][]{2021MNRAS.506.3079K,2024arXiv240319635B}.
Thus, the radio-detected BB~22 is an exception.

In case BB~22 hosts an AGN, we estimated the mass of its central massive black hole (BH). 
Assuming the typical BH mass to stellar mass ratio found in the local Universe \citep[$M_{\rm BH}/M_\star = 0.001$--$0.002$; e.g.,][]{2003ApJ...589L..21M,2013ARA&A..51..511K}, the BH mass of BB~22 is estimated to be around $10^{5.2}$--$10^{5.5} ~ M_\odot$, which would place it near the range of intermediate-mass BHs.
However, the velocity dispersion of BB~22 was found to be $\upsilon_{\rm sys} = 64.4 \pm 5.2$, and based on the M--$\sigma$ relation \citep[][]{2002ApJ...574..740T}, the BH mass is estimated to be $M_{\rm BH} = 10^{6.15 \pm 0.14} ~ M_\odot$.

\subsection{Stellar populations and interstellar medium of blueberry galaxies}
\label{sec:discuss_SP_ages}

The majority of BBs have lower metallicities compared to the expected values based on their $M_\star$ and SFR (Figure \ref{fig:mass_metallicity}).
A possible explanation for this could be the very rapid growth of their overall low stellar mass, driven by strong and recent star-formation bursts.
Considering that BBs are the most highly star-forming dwarf galaxies in the local Universe and also have very low stellar masses ($10^6 < M_\star/M_\odot < 10^9$; Figure \ref{fig:Main_Sequence}), a recent star-formation burst could generate a relatively large fraction of their total number of stars.
Having very recent and strong starbursts could explain the very low metallicities of BBs, considering the time needed for metals to be distributed in the ISM after the star-formation episodes.

Additionally, most of this work's BBs show a stronger extinction compared to what is expected based on their metallicity and are overall below the average relation (Figure \ref{fig:metall_extinction}).
This result also suggests that these galaxies had very recent starburst episodes.
Recent and ongoing star-formation events produce dust.
However, because these galaxies are very young and have a very low stellar mass, dominated by a recent star-formation episode, there have not been many cycles of star formation that would have allowed for the dispersion of metals in the ISM.
Moreover, these sources must have had very recent star-formation episodes since more than a few million years passing in such strong ionizing radiation fields would lead to the destruction of the dust \citep[e.g.,][]{2006ApJ...648..435N,2014A&A...566A.136M}.

The analysis of the BB's IR photometry is complementary to this interpretation.
The red IR colors of these BBs could be explained based on the galaxies having very recent star-formation episodes.
The dominant SPs in BBs are young and massive stars, and their strong UV emission is absorbed by the dust that is present in the birth clouds and the ISM, leading to strong emission in the mid-IR part of the spectrum.
In contrast, the near-IR, which traces the thermal emission of low-mass and long-lived stars, is relatively suppressed due to the absence of a strong population of older stars in BBs. 
Because BBs are galaxies with very low metallicities, we did not expect a relatively high fraction of small dust grains and polycyclic aromatic hydrocarbons (PAHs) to be present (e.g., \citealt{2007ApJ...663..866D,2008ApJ...678..804E}; see also the IR spectrum of the low-metallicity BDG UGCA 219 by \cite{2014ApJS..212...18B}), which could otherwise contribute to the mid-IR emission.
Thus, the progressively redder IR colors of BBs can be mainly attributed to dust that is stochastically heated by the strong radiation field of their young SPs.

It is known that the spectra of dwarf SFGs tend to have increased intensity in the mid-IR wavelengths \citep[e.g.,][]{2014ApJS..212...18B}, while some BDGs can even mimic the IR colors of AGN galaxies \citep[e.g.,][]{2014ApJ...784..113S,2016ApJ...832..119H,2023A&A...679A..76D}.
The sources that are placed in the \cite{2011ApJ...735..112J} and \cite{2018ApJ...858...38S} wedges are mainly BDGs with a very high sSFR, and they are also among the most massive BDGs with similar levels of sSFR. 
This indicates that the presence of a relatively stronger population of older stars, in combination with the ongoing star formation, could be the reason for the softer $W2-W3$ that leads BDGs to mimic the IR colors of AGNs.
The fact that the sSFR of the BDGs correlates more strongly with $W1-W2$ rather than with $W2-W3$ (Figure \ref{fig:IR_colors_sSFR}) indicates that the intensity of the radiation field may also affect the overall shape of their IR spectrum.
In contrast to the more massive BDGs, the very-low-mass BBs tend to have a more stochastic behavior, but most of them are placed in the upper-right part of the diagram and do not enter the \cite{2011ApJ...735..112J} and \cite{2018ApJ...858...38S} wedges.

\subsection{Comparison between blueberry and green pea galaxies}

Blueberry galaxies are considered the lower-mass and nearby analogs of GPs.
Although BBs are generally smaller and less massive, they also show lower metallicities compared to GPs (Figure \ref{fig:mass_metallicity}) and have on average about an order of magnitude higher sSFRs (Figures \ref{fig:Main_Sequence}, \ref{fig:spec_results}).
Moreover, the higher values of EWs of BBs in all the compared emission lines indicate that they are forming stars even faster and that their underlying population of older stars is smaller.
It has been shown that there is a tight correlation between the $\rm EW_{H\alpha}$ and sSFR of galaxies \citep[e.g.,][]{2018MNRAS.477.3014B}.
The EW takes into account the emission-line component over the continuum at the specific wavelength of each line. 
As the continuum traces the stellar emission of low-mass SPs and the emission lines trace the ionizing radiation of the massive stars, the EW is also a tracer of the relative star-forming activity over the stellar component of galaxies.

Comparison of the samples' spectroscopic properties reveals that, on average, BBs show stronger levels of ionization (Figure \ref{fig:spec_results}).
Additionally, most of the spectroscopically observed BBs of this work show $\ion{He}{II} \lambda 4686$ in emission, a high ionization emission line that, in the absence of an AGN, requires the presence of many very massive and young stars. 
The comparison of the IR colors of the two populations is complementary to this conclusion since BBs show on average redder IR colors, mainly in the $\rm W2-W3$ color (Figure \ref{fig:IR_colors}).

Although the GP selection process followed by \cite{2009MNRAS.399.1191C} was not restrictive regarding how blue the galaxies' colors could be, their selected sources were limited to the redshift range $0.112 < z < 0.360$~$\rm (537 \lesssim D/Mpc \lesssim 1978)$.
This led to the exclusion of BB-like sources because BBs, although bluer, are less luminous compared to the more massive GPs.
For instance, the brightest source of this sample, BB~22 (with observed $r=15.6$~mag), would be marginally detectable ($r=21.8$~mag), and the second brightest source, BB~32 (with observed $r=16.8$~mag), would be completely undetectable ($r=23.5$~mag) by SDSS if they were placed at $\sim 1600$~Mpc (z=0.3), a distance representative of most GPs.
Thus, although BBs and GPs are part of the larger population of BDGs, the particular samples have intrinsic differences, with the former being smaller and overall more extreme.

\section{Conclusions}
\label{sec:conclusions}

In order to identify BB galaxies and study their properties probed by infrared and visible light, this study utilized the HECATE catalog, the Pan-STARRS and SDSS photometries, the MPA-JHU spectroscopic analysis, and ALLWISE photometry. 
In summary, this work has done the following:
\begin{itemize}
    \item Identified 48 BBs, 40 of which are first recognized as such from this work. 
    Most are closer compared to the previously known BBs, and 14 are in the south equatorial sky.
    
    \item Reported the distribution of the BBs' physical sizes and showed that BBs have lower metallicities and stronger extinction compared to the expected values when accounting for their $M_\star$ and SFRs.
    
    \item Showed that BBs are at the extreme end of dwarf SFGs.
    Compared to GPs, BBs are on average less massive, and they form stars more intensively, showing more strongly ionized ISM, lower metallicities, intrinsically bluer visible colors, and redder IR colors.
    
    \item Showed that BDGs with higher specific SFRs tend to have redder IR colors, with the correlation being stronger with the $W1-W2$ color.
\end{itemize}

This sample of BBs can be used as a basis for future multi-wavelength studies of some of the most highly star-forming dwarf galaxies of the local Universe. 
Moreover, since BBs and GPs are the best local analogs of the dwarf SFGs of the early Universe, the results of this work provide constraints regarding their optical and IR emission.

\begin{acknowledgements}

The authors thank the anonymous referee for providing comments and suggestions that improved the clarity of this work.
This work was supported by the Czech Science Foundation project No.22-22643S.
This project has received funding from the European Union's Horizon 2020 research and innovation programme under the Marie Sklodowska-Curie RISE action, grant agreements No 691164 (ASTROSTAT), and No 873089 (ASTROSTAT-II).

This research made use of the ``K-corrections calculator'' service available at http://kcor.sai.msu.ru/.
This research has made use of: 
(a) software provided by the CXC in the application packages DS9;
(b) data products from the Wide-field Infrared Survey Explorer (WISE), which is a joint project of the University of California, Los Angeles, and JPL, California Institute of Technology, funded by NASA; 
(c) observations made with the Spitzer Space Telescope, which was operated by JPL, California Institute of Technology under a contract with NASA; 
(d) the NASA/IPAC Extragalactic Database (NED), which is operated by the Jet Propulsion Laboratory (JPL), California Institute of Technology, under contract with NASA; 
(e) the NASA/IPAC Infrared Science Archive (IRSA), which is funded by NASA and operated by the California Institute of Technology; 
Funding for the SDSS and SDSS-II has been provided by the Alfred P. Sloan Foundation, the Participating Institutions, the National Science Foundation, the U.S. Department of Energy, the National Aeronautics and Space Administration, the Japanese Monbukagakusho, the Max Planck Society, and the Higher Education Funding Council for England. The SDSS Web Site is http://www.sdss.org/.

The SDSS is managed by the Astrophysical Research Consortium for the Participating Institutions. The Participating Institutions are the American Museum of Natural History, Astrophysical Institute Potsdam, University of Basel, University of Cambridge, Case Western Reserve University, University of Chicago, Drexel University, Fermilab, the Institute for Advanced Study, the Japan Participation Group, Johns Hopkins University, the Joint Institute for Nuclear Astrophysics, the Kavli Institute for Particle Astrophysics and Cosmology, the Korean Scientist Group, the Chinese Academy of Sciences (LAMOST), Los Alamos National Laboratory, the Max-Planck-Institute for Astronomy (MPIA), the Max-Planck-Institute for Astrophysics (MPA), New Mexico State University, Ohio State University, University of Pittsburgh, University of Portsmouth, Princeton University, the United States Naval Observatory, and the University of Washington.

This research made use of the Digitized Sky Surveys (DSS).
The Digitized Sky Surveys were produced at the Space Telescope Science Institute under U.S. Government grant NAG W-2166. The images of these surveys are based on photographic data obtained using the Oschin Schmidt Telescope on Palomar Mountain and the UK Schmidt Telescope. The plates were processed into the present compressed digital form with the permission of these institutions.
The National Geographic Society - Palomar Observatory Sky Atlas (POSS-I) was made by the California Institute of Technology with grants from the National Geographic Society.
The Second Palomar Observatory Sky Survey (POSS-II) was made by the California Institute of Technology with funds from the National Science Foundation, the National Geographic Society, the Sloan Foundation, the Samuel Oschin Foundation, and the Eastman Kodak Corporation.
The Oschin Schmidt Telescope is operated by the California Institute of Technology and Palomar Observatory.
The UK Schmidt Telescope was operated by the Royal Observatory Edinburgh, with funding from the UK Science and Engineering Research Council (later the UK Particle Physics and Astronomy Research Council), until 1988 June, and thereafter by the Anglo-Australian Observatory. The blue plates of the southern Sky Atlas and its Equatorial Extension (together known as the SERC-J), as well as the Equatorial Red (ER), and the Second Epoch [red] Survey (SES) were all taken with the UK Schmidt.
All data are subject to the copyright given in the copyright summary. Copyright information specific to individual plates is provided in the downloaded FITS headers.

Supplemental funding for sky-survey work at the ST ScI is provided by the European Southern Observatory. 


\end{acknowledgements}

\bibliographystyle{aa} 
\bibliography{citations.bib} 

\begin{appendix}
\FloatBarrier
\section{The selection algorithm}
\label{sec:Appendix_A}

\numberwithin{table}{section}
\setcounter{table}{1}
\renewcommand{\theequation}{A\arabic{table}}

The following Equation \ref{eq:Selection} shows the algorithm used for the selection of BB galaxies (as described in Section \ref{sec:criteria}):

\begin{equation}
\begin{aligned}
    {\rm Include:} \\ 
    {\rm if} ~ 
    \Bigg( (g-r)_{\rm cMod,err} < 0.2 ~ {\rm and } -2 < (g-r)_{\rm cMod} < -0.4 ~ {\rm and } \\
    (r-i)_{\rm cMod,err} < 0.2 ~ {\rm and } -2 < (r-i)_{\rm cMod} < 1.1 ~ {\rm and } \\
    (g-i)_{\rm cMod,err} < 0.2 ~ {\rm and } -2 < (g-i)_{\rm cMod} < -0.4 ~ {\rm and } \\
    (g-u)_{\rm cMod,err} < 0.2 ~ {\rm and } -2 < (g-u)_{\rm cMod} < -0.2 ~ {\rm and } \\
    (g-r)_{\rm Petro,err} < 0.2 ~ {\rm and } -2 < (g-r)_{\rm Petro} < -0.4 ~ {\rm and } \\
    (r-i)_{\rm Petro,err} < 0.2 ~ {\rm and } -2 < (r-i)_{\rm Petro} < 1.1 ~ {\rm and } \\
    (g-i)_{\rm Petro,err} < 0.2 ~ {\rm and } -2 < (g-i)_{\rm Petro} < -0.4 ~ {\rm and } \\
    (g-u)_{\rm Petro,err} < 0.2 ~ {\rm and } -2 < (g-u)_{\rm Petro} < -0.2
    \Bigg)\\
    {\rm if} ~ 
    \Bigg( (g-r)_{\rm Kron,err} < 0.2 ~ {\rm and } -2 < (g-r)_{\rm Kron} < -0.25 ~ {\rm and } \\
    (r-i)_{\rm Kron,err} < 0.2 ~ {\rm and } -2 < (r-i)_{\rm Kron} < 1.25 ~ {\rm and } \\
    (g-i)_{\rm Kron,err} < 0.2 ~ {\rm and } -2 < (g-i)_{\rm Kron} < -0.25 ~ {\rm and } \\
    (g-u)_{\rm Kron,err} < 0.2 ~ {\rm and } -2 < (g-u)_{\rm Kron} < -0.05 ~ {\rm and } \\
    (g-r)_{\rm PSF,err} < 0.2 ~ {\rm and } -2 < (g-r)_{\rm PSF} < -0.25 ~ {\rm and } \\
    (r-i)_{\rm PSF,err} < 0.2 ~ {\rm and } -2 < (r-i)_{\rm PSF} < 1.25 ~ {\rm and } \\
    (g-i)_{\rm PSF,err} < 0.2 ~ {\rm and } -2 < (g-i)_{\rm PSF} < -0.25 ~ {\rm and } \\
    (g-u)_{\rm PSF,err} < 0.2 ~ {\rm and } -2 < (g-u)_{\rm PSF} < -0.05
    \Bigg) \\
    {\rm Exclude:}\\ 
    {\rm if} ~ 
    \Bigg( (g-r)_{\rm mean} > -0.25 ~ {\rm or } \\
    (r-i)_{\rm mean} > 1.25 ~ {\rm or } \\
    (g-i)_{\rm mean} > -0.25 ~ {\rm or } \\
    (g-r)_{\rm mean} > -0.05 \Bigg). \\
\label{eq:Selection}
\end{aligned}
\end{equation}

\section{The spectral energy distribution fitting process}
\label{sec:Appendix_B}

\begin{figure}
    \centering
    \includegraphics[width=\columnwidth]{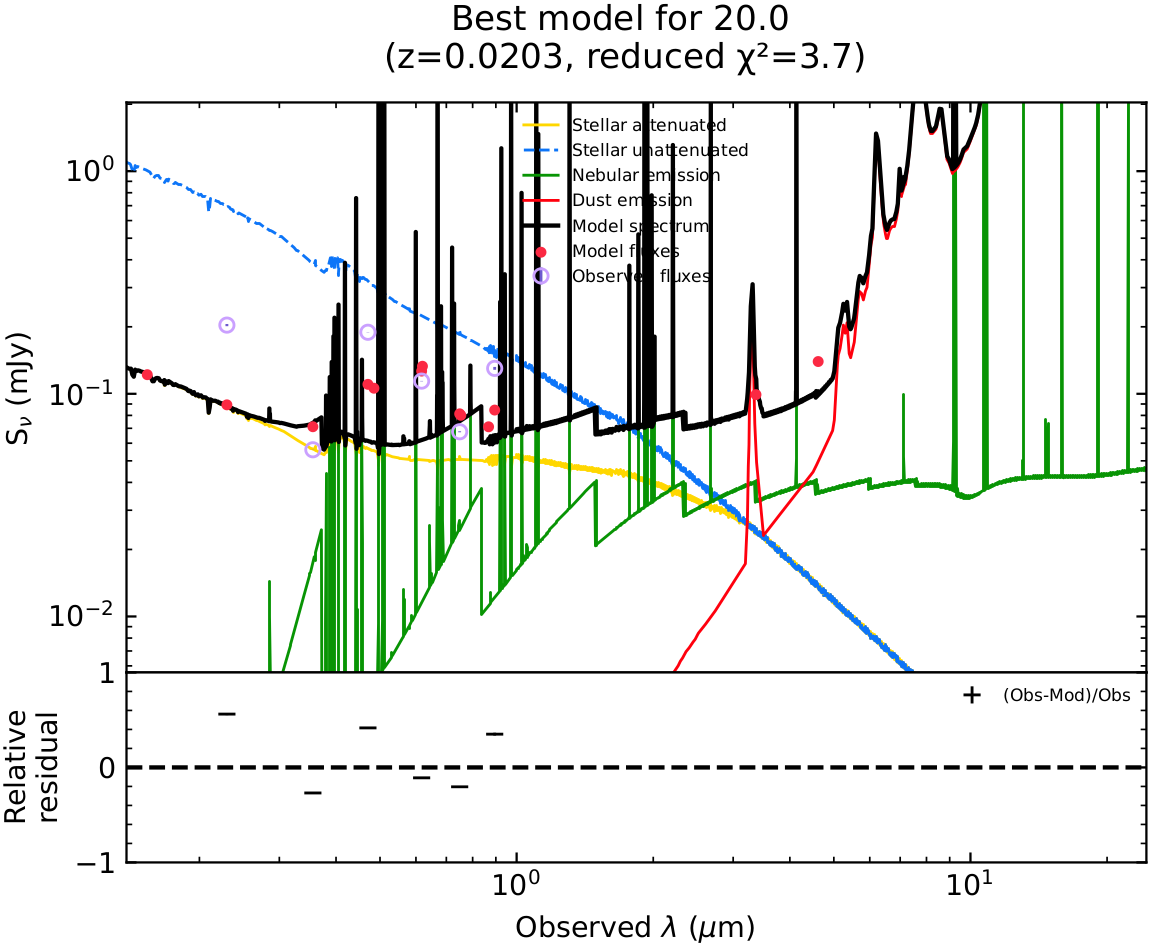}
    \includegraphics[width=\columnwidth]{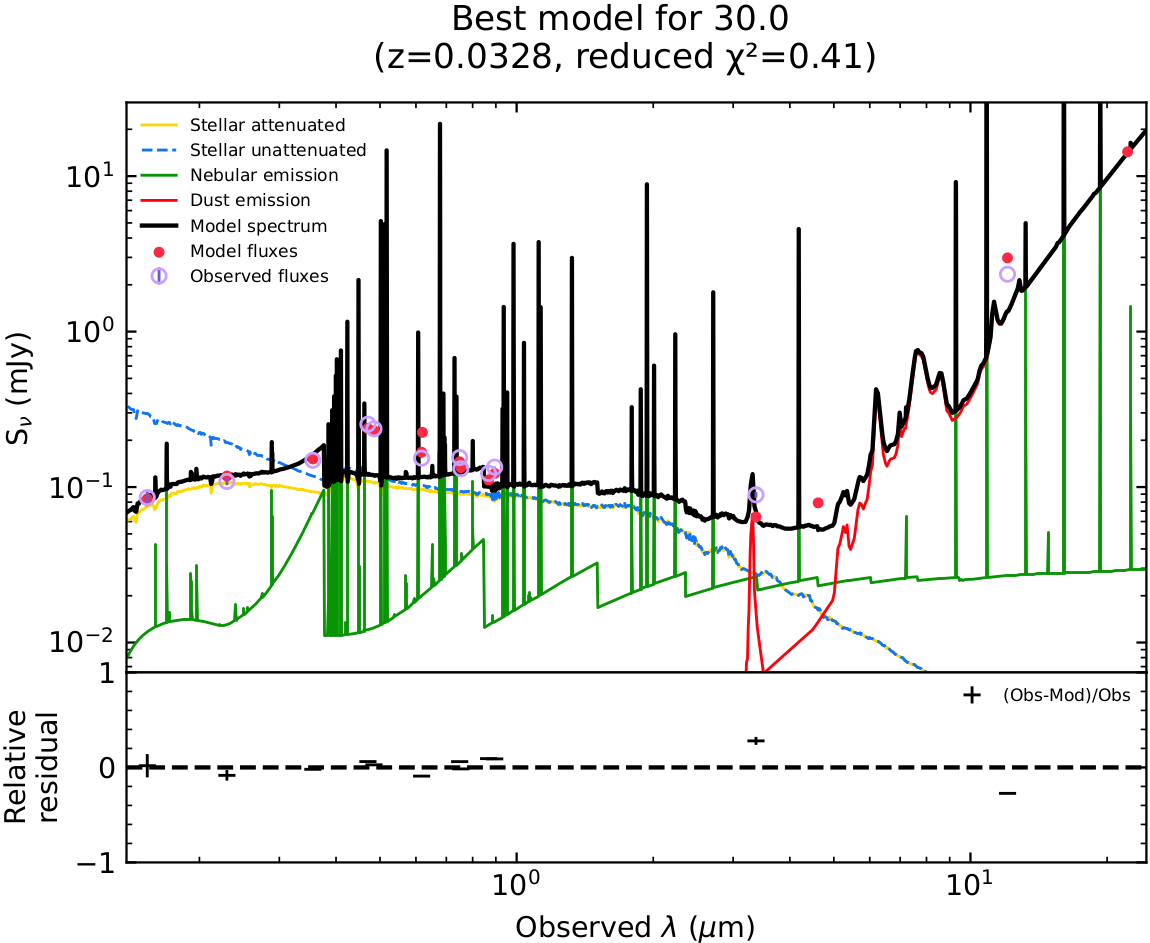}
    \caption{Spectral energy distributions and \texttt{CIGALE} fits for BB~20 (top panel) and BB~30 (bottom panel).
    The stellar attenuated and unattenuated continuum is shown with yellow and blue dashed lines, respectively.
    The green, red, and black lines show the nebular, dust, and total emission, respectively.
    The observed fluxes are shown with purple open circles, and the estimated fluxes are shown with red circles.}
    \label{fig:CIGALE_fits}
\end{figure}

\begin{figure*}
    \centering
    \includegraphics[width=0.93\textwidth]{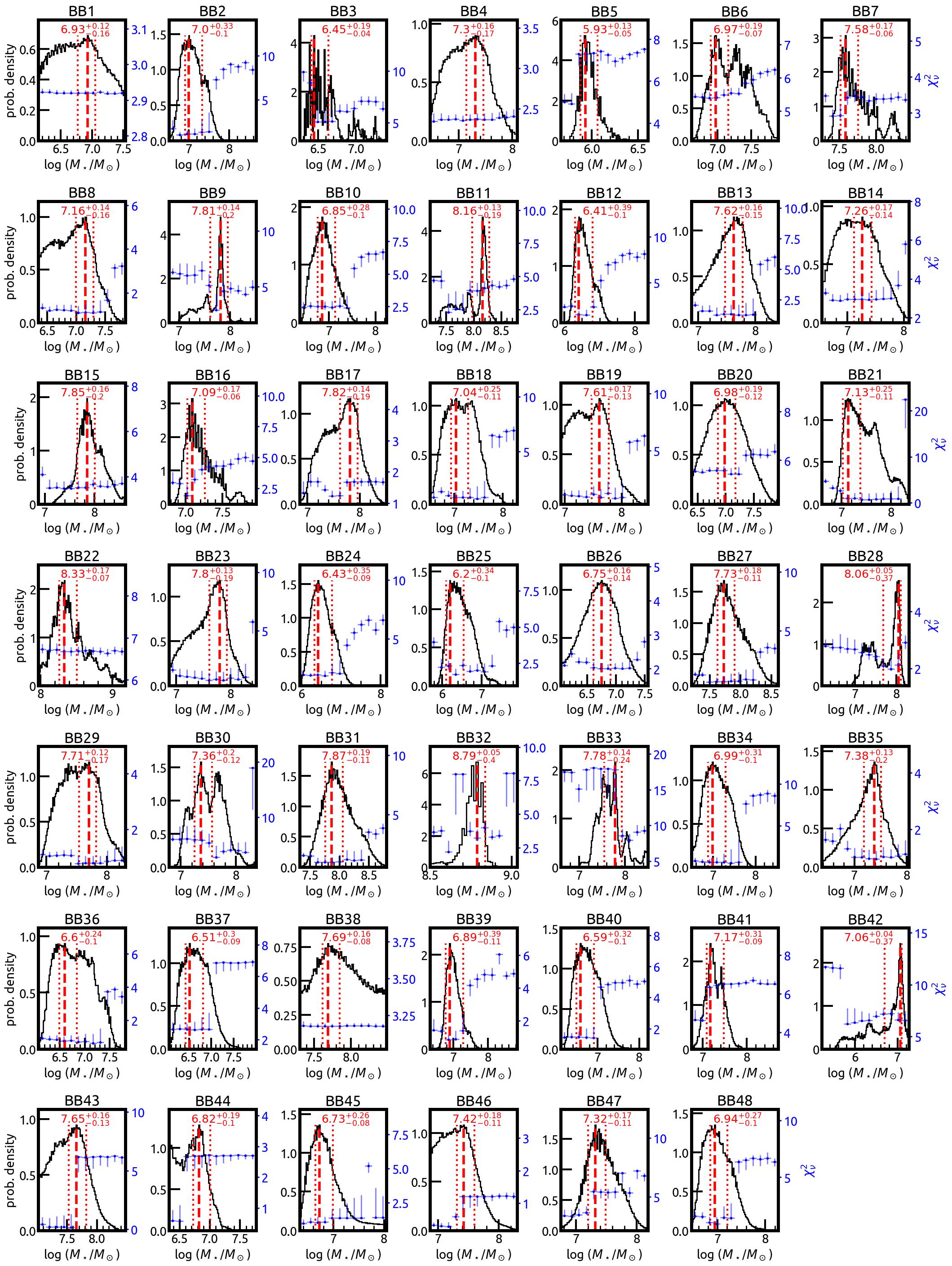}
    \caption{Stellar mass $\chi^{2}_{\nu}$-weighted probability densities of all \texttt{CIGALE} models  (black line). 
    Each panel represents the modeling of an individual BB.
    The distribution modes and CIs are shown with vertical dashed and dotted red lines, respectively.
    The logarithm of the estimated $M_\star$ is shown at the top of each panel with red fonts.
    The blue error bars represent the bin modes and CIs of the model's $\chi^{2}_\nu$ at the plotted $M_\star$ range divided into 12 bins, while the right y-axis corresponds to the $\chi^{2}_\nu$ values.}
    \label{fig:CIGALE_Mstar}
\end{figure*}

\begin{figure*}
    \centering
    \includegraphics[width=0.93\textwidth]{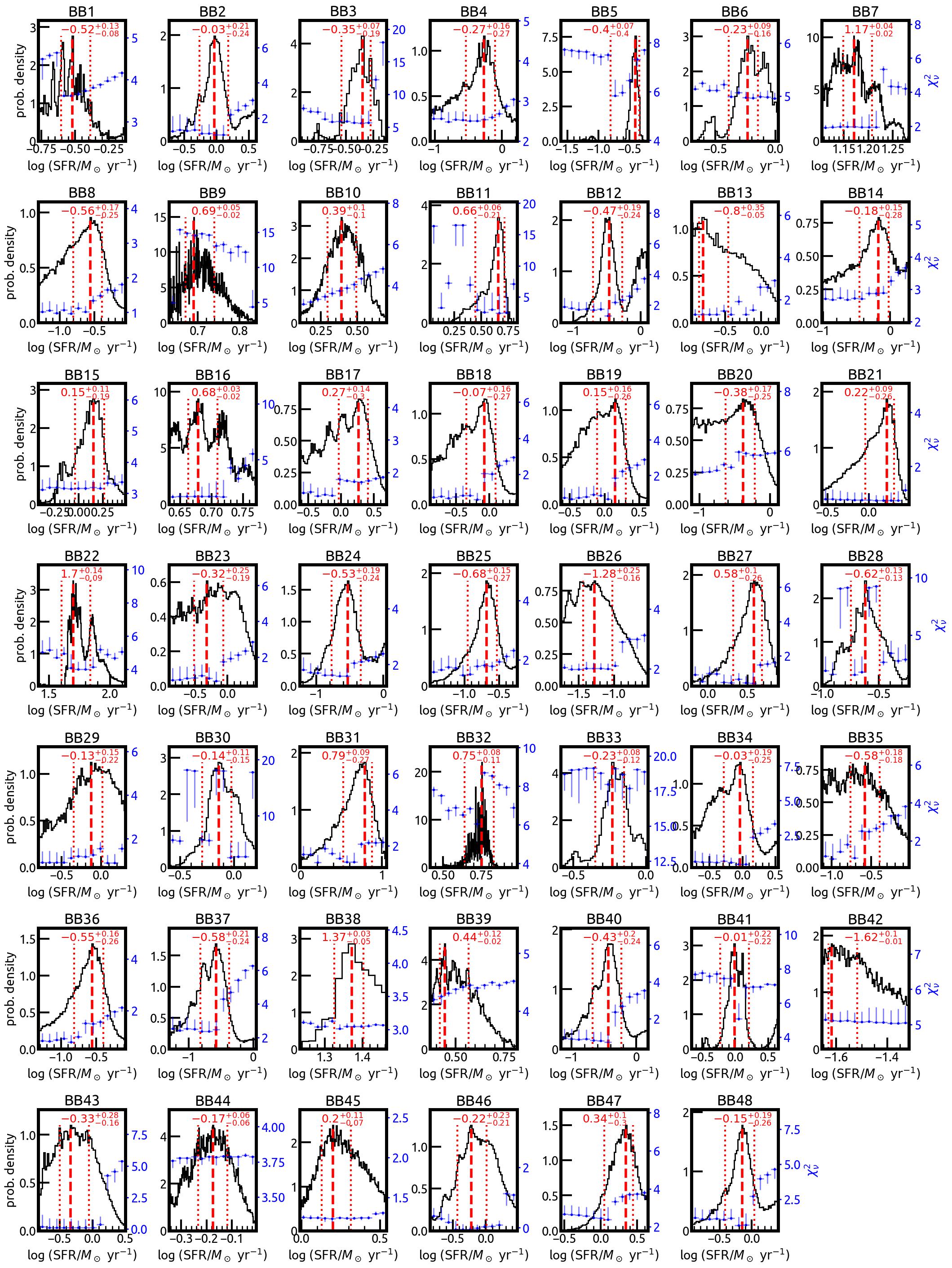}
    \caption{Star-formation rate $\chi^{2}_{\nu}$-weighted probability densities of all \texttt{CIGALE} models (black line). 
    Each panel shows the modeling of an individual BB.
    The modes and CIs are shown with vertical dashed and dotted red lines, respectively.
    The estimated SFR is shown at the top of each panel in red.
    The blue error bars represent the bin modes and CIs of the model's $\chi^{2}_\nu$ at the plotted SFR range divided into 12 bins, while the right y-axis corresponds to the $\chi^{2}_\nu$ values.}
    \label{fig:CIGALE_SFR}
\end{figure*}

Figure \ref{fig:CIGALE_fits} shows two representative SED fits performed with \texttt{CIGALE}, namely the worst (BB~20) and one of the best (BB~30), based on $\chi^{2}_{\nu}$. 
Figures \ref{fig:CIGALE_Mstar} and \ref{fig:CIGALE_SFR} show the $\chi^{2}_{\nu}$-weighted probability densities, accounting for all fitted models, of the $M_\star$ and SFR, respectively, for the 48 BBs of this work.
The SFR and $M_\star$ quantities were estimated based on the modes and 68\% CIs of these distributions.

\end{appendix}

\end{document}